\documentclass[12pt]{article}

\usepackage[utf8]{inputenc}
\usepackage{textcomp}
\usepackage{chetnew}
\usepackage{tikz}
\usepackage{amssymb,amsfonts,mathrsfs,dsfont,yfonts,bbm}\usetikzlibrary{calc}
\usepackage{xcolor}
\usepackage{braket}

\newcommand{\CH}{\mathcal{H}}
\newcommand{\CD}{\mathcal{D}}

\newcommand{\CQ}{\mathcal{Q}}
\newcommand{\CB}{\mathcal{B}}
\newcommand{\CC}{\mathcal{C}}
\newcommand{\CO}{\mathcal{O}}
\newcommand{\CT}{\mathcal{T}}
\newcommand{\CI}{\mathcal{I}}
\newcommand{\CN}{\mathcal{N}}
\newcommand{\CS}{\mathcal{S}}
\newcommand{\CM}{\mathcal{M}}

\newcommand*{\boxcoloro}{orange}
\makeatletter
\newcommand{\boxedo}[1]{\textcolor{\boxcoloro}{%
\tikz[baseline={([yshift=-1ex]current bounding box.center)}] \node [rectangle, minimum width=1ex,rounded corners,draw] {\normalcolor\m@th$\displaystyle#1$};}}
 \makeatother
\newcommand*{\boxcolorr}{red}
\makeatletter
\newcommand{\boxedr}[1]{\textcolor{\boxcolorr}{%
\tikz[baseline={([yshift=-1ex]current bounding box.center)}] \node [rectangle, minimum width=1ex,rounded corners,draw] {\normalcolor\m@th$\displaystyle#1$};}}
 \makeatother
\newcommand*{\boxcolorb}{blue}
\makeatletter
\newcommand{\boxedb}[1]{\textcolor{\boxcolorb}{%
\tikz[baseline={([yshift=-1ex]current bounding box.center)}] \node [rectangle, minimum width=1ex,rounded corners,draw] {\normalcolor\m@th$\displaystyle#1$};}}
 \makeatother
\newcommand*{\boxcolorg}{green}
\makeatletter
\newcommand{\boxedg}[1]{\textcolor{\boxcolorg}{%
\tikz[baseline={([yshift=-1ex]current bounding box.center)}] \node [rectangle, minimum width=1ex,rounded corners,draw] {\normalcolor\m@th$\displaystyle#1$};}}
 \makeatother
 \newcommand*{\boxcolorp}{purple}
\makeatletter
\newcommand{\boxedp}[1]{\textcolor{\boxcolorp}{%
\tikz[baseline={([yshift=-1ex]current bounding box.center)}] \node [rectangle, minimum width=1ex,rounded corners,draw] {\normalcolor\m@th$\displaystyle#1$};}}
 \makeatother
  \newcommand*{\boxcolorc}{cyan}
\makeatletter
\newcommand{\boxedc}[1]{\textcolor{\boxcolorc}{%
\tikz[baseline={([yshift=-1ex]current bounding box.center)}] \node [rectangle, minimum width=1ex,rounded corners,draw] {\normalcolor\m@th$\displaystyle#1$};}}
 \makeatother
  \newcommand*{\boxcolory}{yellow}
\makeatletter
\newcommand{\boxedy}[1]{\textcolor{\boxcolory}{%
\tikz[baseline={([yshift=-1ex]current bounding box.center)}] \node [rectangle, minimum width=1ex,rounded corners,draw] {\normalcolor\m@th$\displaystyle#1$};}}
 \makeatother

\begin{document}
\preprint{QMUL-PH-17-XX}

\title{$\CN=2$ $S$-duality Revisited}

\author{Matthew Buican$^{\diamondsuit, 1}$,  Zoltan Laczko$^{\clubsuit, 1}$, and Takahiro Nishinaka$^{\heartsuit, 2}$}

\affiliation{\smallskip$^1$ CRST and School of Physics and Astronomy\\
Queen Mary University of London, London E1 4NS, UK\\  \smallskip$^2$ Department of Physical Sciences, College of Science and Engineering \\ Ritsumeikan University, Shiga 525-8577, Japan\emails{$^{\diamondsuit}$m.buican@qmul.ac.uk, $^{\clubsuit}$ z.b.laczko@qmul.ac.uk,$^{\heartsuit}$nishinak@fc.ritsumei.ac.jp}}

\abstract{Using the chiral algebra bootstrap, we revisit the simplest Argyres-Douglas (AD) generalization of Argyres-Seiberg $S$-duality. We argue that the exotic AD superconformal field theory (SCFT), $\CT_{3,{3\over2}}$, emerging in this duality splits into a free piece and an interacting piece, $\CT_X$, even though this factorization seems invisible in the Seiberg-Witten (SW) curve derived from the corresponding M5-brane construction. Without a Lagrangian, an associated topological field theory, a BPS spectrum, or even an SW curve, we nonetheless obtain exact information about $\CT_X$ by bootstrapping its chiral algebra, $\chi(\CT_X)$, and finding the corresponding vacuum character in terms of Affine Kac-Moody characters. By a standard 4D/2D correspondence, this result gives us the Schur index for $\CT_X$ and, by studying this quantity in the limit of small $S^1$, we make contact with a proposed $S^1$ reduction. Along the way, we discuss various properties of $\CT_X$: as an $\CN=1$ theory, it has flavor symmetry $SU(3)\times SU(2)\times U(1)$, the central charge of $\chi(\CT_X)$ matches the central charge of the $bc$ ghosts in bosonic string theory, and its global $SU(2)$ symmetry has a Witten anomaly. This anomaly does not prevent us from building conformal manifolds out of arbitrary numbers of $\CT_X$ theories (giving us a surprisingly close AD relative of Gaiotto's $T_N$ theories), but it does lead to some open questions in the context of the chiral algebra / 4D $\CN=2$ SCFT correspondence.}

\date{June 2017}

\setcounter{tocdepth}{2}

\maketitle
\toc

\newsec{Introduction}
Four-dimensional (4D) superconformal field theories (SCFTs) often admit exactly marginal deformations (the spaces of these deformations are typically called \lq\lq conformal manifolds"). In the context of theories with $\CN\ge2$ supersymmetry (SUSY), one can easily obtain examples with exactly marginal deformations by coupling a gauge multiplet to precisely enough matter so that the one-loop beta function vanishes. A canonical example of this phenomenon occurs in $su(N)$ $\CN=4$ Super Yang-Mills (SYM). At the level of the Lie algebra and the local operators, this theory is self-dual:\footnote{See the recent analysis in \cite{Aharony:2013hda} for a discussion of subtleties at the level of the gauge group and the line operators.} as we vary the exactly marginal gauge coupling, $\tau$, towards a strong-coupling cusp on the conformal manifold, an $S$-dual weakly coupled $su(n)$ $\CN=4$ SYM theory emerges. A similar story holds in $su(2)$ $\CN=2$ gauge theory with four fundamental flavors \cite{Seiberg:1994aj}.

On the other hand, the $S$-duality in $su(3)$ $\CN=2$ gauge theory with six fundamental flavors is dramatically different \cite{Argyres:2007cn}. As one takes the gauge coupling to infinity, Argyres and Seiberg found that, instead of getting a weakly coupled $S$-dual description in terms of another $su(3)$ gauge theory with fundamental matter, one instead finds a dual consisting of an $su(2)$ theory coupled to a doublet of hypermultiplets and an $su(2)\subset\mathfrak{e}_6$ factor of the global symmetry of the Minahan-Nemeschansky $E_6$ SCFT \cite{Minahan:1996fg}.

The message of \cite{Argyres:2007cn} is clear: sometimes, starting from vanilla building blocks, the \lq\lq matter" that appears via $\CN=2$ $S$-duality is not standard matter (i.e., hypermultiplets) but is instead a strongly coupled isolated SCFT\footnote{By \lq\lq isolated," we mean a theory that lacks an exactly marginal deformation.} whose global symmetry (or a proper subgroup thereof) is weakly gauged.\footnote{The corresponding contribution to the beta function---the current two point function coefficient, $k$---is often exactly computable since it is given by a contact term in the correlator of the superconformal $U(1)_R$ current with two flavor currents.} Moreover, $S$-duality can be a machine for generating exotic isolated theories.

This latter point was driven home in \cite{Gaiotto:2009we}. Indeed, Gaiotto generalized \cite{Argyres:2007cn} to higher-rank gauge theories and, in the process, found an infinite number of new isolated SCFTs---the so-called $T_N$ theories---at strong-coupling cusps on the resulting conformal manifolds.\footnote{The $T_3$ case is just the $E_6$ SCFT of \cite{Minahan:1996fg}, and the $T_2$ case is eight free half-hypermultiplets. However, the $T_N$ SCFTs with $N\ge4$ are new isolated theories.} Since a $T_N$ theory has $SU(N)^3$ global symmetry\footnote{The $T_3$ case has an enhanced $E_6\supset SU(3)^3$ global symmetry, but the discussion below applies to this theory as well. A similar discussion holds for the $T_2$ theory, which has $Sp(4)\supset SU(2)^3$ global symmetry.} and the following $SU(N)$ current two-point function (and hence 1-loop beta function contribution upon gauging) for each such factor
\begin{equation}\label{TNk}
k_{SU(N)_{i}}^{T_N}=2N~,\ \ \ i=1,2,3~,
\end{equation}
one can always find a non-trivial conformal manifold by taking two $T_N$ theories and gauging a diagonal $SU(N)$. Indeed, the contributions from the $T_N$ theories in \eqref{TNk} cancel those of the $SU(N)$ gauge fields
\begin{equation}\label{1loop}
\beta^{\rm 1-loop}_{SU(N)}=-4N+2N+2N=0~.
\end{equation}
One can then proceed to construct a conformal manifold consisting only of arbitrarily many $T_N$ theories and conformal gauge fields.

While the above set of theories is quite vast, the $T_N$ theories (and their cousins) are somewhat special: their $\CN=2$ chiral primaries have integer scaling dimensions.\footnote{By $\CN=2$ chiral primaries, we mean superconformal primaries that are annihilated by all the anti-chiral Poincar\'e supercharges of $\CN=2$ SUSY.} The underlying reason is that these theories emerge in a duality with a Lagrangian theory.\footnote{By the rules of \cite{Dolan:2002zh}, $\CN=2$ chiral operators cannot disappear from the spectrum or, by the discussion in \cite{Papadodimas:2009eu}, have their dimensions renormalized as we vary $\tau$, so the $T_N$ $\CN=2$ chiral ring generators must correspond to some subset of the gauge Casimirs of a Lagrangian theory.}  On the other hand, the most generally allowed values for the scaling dimensions, $\Delta_i$, of $\CN=2$ chiral operators are widely believed to be $\Delta_i\in\mathbb{Q}$, and non-integer rational values are indeed realized in so-called Argyres-Douglas (AD) theories \cite{Argyres:1995jj,Argyres:1995xn,Xie:2012hs}.\footnote{We define any $\CN=2$ SCFT with non-integer scaling dimension chiral primaries to be of AD type.} These theories cannot emerge in an $\CN=2$ $S$-duality with a Lagrangian theory.

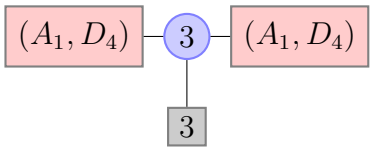
\begin{figure}
\begin{center}
\vskip .5cm
\begin{tikzpicture}[place/.style={circle,draw=blue!50,fill=blue!20,thick,inner sep=0pt,minimum size=6mm},transition/.style={rectangle,draw=black!50,fill=black!20,thick,inner sep=0pt,minimum size=5mm},transition2/.style={rectangle,draw=black!50,fill=red!20,thick,inner sep=0pt,minimum size=8mm},auto]
\node[transition2] (1) at (1.5,0) {$\ (A_1, D_4)\ $};
\node[place] (2) at (3,0) [shape=circle] {$3$} edge [-] node[auto]{} (1);
\node[transition2] (3) at (4.5,0)  {$\ (A_1, D_4)\ $} edge [-] node[auto]{} (2);
\node[transition] (9) at (3,-1.2) {$3$} edge[-] (2);
\end{tikzpicture}
\caption{The quiver diagram describing the simplest (i.e., lowest rank) AD generalization of Argyres-Seiberg duality in the $SU(3)$ duality frame. The total flavor symmetry is $U(3)$. In \cite{Buican:2014hfa}, this theory was called the \lq\lq $\CT_{3,2,{3\over2},{3\over2}}$" SCFT.}
\label{quiver1}
\end{center}
\end{figure}

Motivated by a desire to understand $\CN=2$ $S$-duality more broadly, it is then natural to ask what is the minimal (which we will define to be lowest rank\footnote{By rank, we mean the complex dimension of the Coulomb branch.}) AD generalization of Argyres-Seiberg (i.e., non self-similar) duality \cite{Buican:2014hfa}. Since the starting point cannot be a Lagrangian theory, one must engineer such a conformal manifold from a weakly coupled gauging of a global symmetry of a collection of AD building blocks (potentially with additional hypermultiplets). An answer, using general consistency conditions and the class $\CS$ Argyres-Douglas theories in \cite{Xie:2012hs}, was given in \cite{Buican:2014hfa} and is reproduced in Fig. \ref{quiver1} (there, this theory was referred to as the \lq\lq$\CT_{3,2,{3\over2},{3\over2}}$" SCFT). This theory is constructed by gauging the diagonal $SU(3)$ symmetry of three fundamental flavors and a pair of $(A_1, D_4)$ SCFTs (the $(A_1, D_4)$ theory, originally discussed in \cite{Argyres:1995xn}, has $SU(3)$ flavor symmetry and a single $\CN=2$ chiral ring generator of dimension $3/2$). The resulting global symmetry is $U(3)$ and is furnished by the three fundamental flavors.

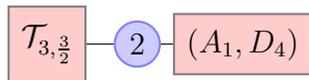
\begin{figure}
\begin{center}
\vskip .5cm
\begin{tikzpicture}[place/.style={circle,draw=blue!50,fill=blue!20,thick,inner sep=0pt,minimum size=6mm},transition/.style={rectangle,draw=black!50,fill=black!20,thick,inner sep=0pt,minimum size=5mm},transition2/.style={rectangle,draw=black!50,fill=red!20,thick,inner sep=0pt,minimum size=8mm},transition3/.style={rectangle,draw=black!50,fill=red!20,thick,inner sep=0pt,minimum size=10mm},auto]
\node[transition3] (1) at (1.8,0) {$\ \CT_{3,{3\over2}}\ $};
\node[place] (2) at (3,0) [shape=circle] {$2$} edge [-] node[auto]{} (1);
\node[transition2] (3) at (4.4,0)  {$\ (A_1, D_4)\ $} edge [-] node[auto]{} (2);
\end{tikzpicture}
\caption{The quiver diagram describing the theory dual to the one in Fig. \ref{quiver1}. The $SU(3)\subset U(3)$ symmetry is furnished by the $\CT_{3,{3\over2}}$ theory while the $U(1)\subset U(3)$ symmetry is furnished by the $(A_1, D_4)$ SCFT. In \cite{Buican:2014hfa}, this theory was called the \lq\lq $\CT_{3,2,{3\over2},{3\over2}}$" SCFT.}
\label{quiver2}
\end{center}
\end{figure}

The $S$-dual frame of this theory is given in Fig. \ref{quiver2} and consists of an $SU(2)$ gauge theory coupled to an $(A_1, D_4)$ factor and a more exotic AD theory called the $\CT_{3,{3\over 2}}$ SCFT \cite{Buican:2014hfa} which has flavor symmetry $G\supset SU(3)\times SU(2)$.\footnote{This latter theory first appeared in the classification of \cite{Xie:2012hs} (using the nomenclature of this paper, $\CT_{3,{3\over2}}$ is a  \lq\lq Type III" theory with Young diagrams $[2,2,2],[2,2,2],[2,2,1,1]$).} Therefore, in rough analogy with Argyres-Seiberg duality, the strongly coupled $(A_1, D_4)$ theory plays the role of the hypermultiplets on the $SU(2)$ side of the duality and the $\CT_{3,{3\over2}}$ theory plays the role of the $E_6=T_3$ theory.

However, upon closer inspection, the analogy with Argyres-Seiberg duality seems to break down. Indeed, the anomalies of the $\CT_{3,{3\over2}}$ theory were computed in \cite{Buican:2014hfa} and found to be
\begin{equation}\label{T332anom}
k_{SU(2)}^{\CT_{3,{3\over2}}}=5~, \ \ \ k_{SU(3)}^{\CT_{3,{3\over2}}}=6~, \ \ \ c^{\CT_{3,{3\over2}}}={9\over4}~, \ \ \ a^{\CT_{3,{3\over2}}}=2~.
\end{equation}
Using these symmetries, one cannot construct conformal manifolds built only out of arbitrary numbers of $\CT_{3,{3\over2}}$ SCFTs and conformal gauge fields. The reason is that the contribution to the $SU(2)$ beta function in \eqref{T332anom} is too large and the required $SU(2)$ gauging would be infrared (IR) free. This state of affairs is quite unlike the $E_6=T_3$ case described above, where an arbitrary number of such theories can be concatenated by gauging enough diagonal symmetries.

Still, there are some puzzles in the above picture. To begin with, the flavor symmetry group of the $\CT_{3,{3\over2}}$ theory is not obvious. One standard way to find such symmetries for SCFTs that, like the $\CT_{3,{3\over2}}$ theory, can be derived from M5-branes wrapping a (punctured) Riemann surface, $\CC$, (so-called class $\CS$ theories) is to construct the Hitchin system corresponding to the theory \cite{Xie:2012hs,Gaiotto:2009hg}. In particular, the Hitchin system has a meromorphic 1-form, $\varphi(z)dz$, with singularities at the punctures of $\CC$. In the case of the $\CT_{3,{3\over2}}$ SCFT, one can construct the corresponding $\varphi$ using the methods in \cite{Xie:2012hs}
\begin{equation}\label{Hfield}
\varphi(z)=z M_1+M_2+{1\over z}M_3+\CO(z^{-2})~,
\end{equation}
where we have expanded around a third-order pole at $z=\infty$ ($\varphi$ is non-singular at all other points $z\in\CC=\mathbb{CP}^1$), and the $M_i$ are the following diagonal traceless matrices
\begin{eqnarray}\label{matrices}
M_1&=&{\rm diag}\left(\tilde a_1, \tilde a_1,\tilde a_2,\tilde a_2,\tilde a_3,\tilde a_3\right)~, \ \ \ M_2={\rm diag}\left(\tilde b_1, \tilde b_1,\tilde b_2,\tilde b_2,\tilde b_3,\tilde b_3\right)~, \cr M_3&=&{\rm diag}\left(\tilde m_1, \tilde m_1,\tilde m_2,\tilde m_2,\tilde m_3,\tilde m_4\right)~.
\end{eqnarray}
The flavor symmetries are then read off by studying the independent parameters appearing as coefficients of the simple pole, i.e., the entries of $M_3$.\footnote{This data gives us the Cartans of the flavor symmetry. By studying various limits of the Hitchin system, we can often identify the full flavor symmetry by matching onto Hitchin sub-systems with known flavor symmetries.} This traceless matrix has three degrees of freedom which correspond to the Cartans of $SU(3)\times SU(2)$. Therefore, according to this description, $G_{\CT_{3,{3\over2}}}=SU(3)\times SU(2)$. One reaches the same conclusion by constructing the Seiberg-Witten (SW) curve from this description via the spectral curve, $\det\left(xdz-\varphi(z)dz\right)=0$, and looking at the mass parameters (i.e., the simple poles in the SW 1-form, $\lambda=xdz$).

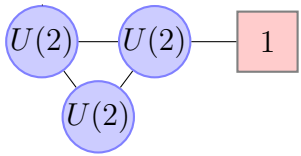
\begin{figure}
\begin{center}
\vskip .5cm
\begin{tikzpicture}[place/.style={circle,draw=blue!50,fill=blue!20,thick,inner sep=0pt,minimum size=6mm},transition/.style={rectangle,draw=black!50,fill=black!20,thick,inner sep=0pt,minimum size=5mm},transition2/.style={rectangle,draw=black!50,fill=red!20,thick,inner sep=0pt,minimum size=8mm},auto]
\node[place] (2) at (.5,0) [shape=circle] {$U(2)$} edge [-] node[auto]{} (2);
\node[place] (3) at (2.,0) [shape=circle] {$U(2)$} edge [-] node[auto]{} (2);
\node[place] (4) at (1.25,-1) [shape=circle] {$U(2)$} edge[-] (2) edge[-] (3);
\node[transition2] (1) at (3.5,0) {$1$} edge[-] (3);
\end{tikzpicture}
\caption{The quiver diagram describing the mirror of the $S^1$ reduction of $\CT_{3,{3\over2}}$.}
\label{quiver3}
\end{center}
\end{figure}

On the other hand, one often computes flavor symmetries of strongly interacting 4D $\CN=2$ theories by taking their $S^1$ reductions and studying the mirror theory (which may sometimes be described by a Lagrangian that flows to the same 3D $\CN=4$ SCFT). Now, the $\CT_{3,{3\over2}}$ theory has a proposed Lagrangian mirror for its $S^1$ reduction given in Fig. \ref{quiver3} (following the rules in \cite{Xie:2012hs}) that predicts flavor symmetry $G^{3d}_{\CT_{3,{3\over2}}}= SU(3)\times SU(2)^2$. Indeed, IR dimension-one monopole operators in this theory describe the enhancement of the manifest $U(1)^3$ topological symmetry to $SU(3)\times SU(2)^2$ \cite{Buican:2014hfa}. In particular, there is a free monopole operator in the IR that gives rise to an additional $SU(2)$ factor.\footnote{This result is somewhat counterintuitive since the rules derived in \cite{Gaiotto:2008ak} for the case of linear quivers suggest that the presence of a free monopole operator can be detected by looking at each gauge node in the quiver and counting the number of local flavors. If this number reaches a certain threshold, then the theory produces a free monopole after one turns on the corresponding gauge coupling(s) and flows to the IR (the theory is then referred to as \lq\lq ugly" in the nomenclature of \cite{Gaiotto:2008ak}). However, it is straightforward to check that the quiver in Fig. \ref{quiver3} should have no free monopoles by these tests and no accidental superconformal $R$ symmetries. The resolution to this puzzle is that the free monopole depends on the global topology of the quiver---it has non-trivial flux through each gauge node---and so the linear quiver tests of \cite{Gaiotto:2008ak} do not apply. \label{topology}} By mirror symmetry \cite{Intriligator:1996ex}, one expects, upon performing an $S^1$ reduction, the enhancement of $G_{\CT_{3,{3\over2}}}\to SU(3)\times SU(2)^2$ with a decoupled hypermultiplet. 

\begin{figure}
\begin{center}
\vskip .5cm
\begin{tikzpicture}[place/.style={circle,draw=blue!50,fill=blue!20,thick,inner sep=0pt,minimum size=6mm},transition/.style={rectangle,draw=black!50,fill=black!20,thick,inner sep=0pt,minimum size=5mm},transition2/.style={rectangle,draw=black!50,fill=red!20,thick,inner sep=0pt,minimum size=8mm},transition3/.style={rectangle,draw=black!50,fill=red!20,thick,inner sep=0pt,minimum size=10mm},auto]
\node[transition3] (1) at (1.8,0) {$\;\CT_{3,{3\over2}}\;$};
\node[] at (3,0) {$=$};
\node[transition2] (1) at (4.1,0){$1\ $};
\node[] at (5,0) {$\oplus$};
\node[transition2] (1) at (5.9,0){$\ \CT_X\ $};
\end{tikzpicture}
\caption{The factorized form of the $\CT_{3,{3\over2}}$ SCFT into a decoupled free hypermultiplet and the interacting $\CT_X$ SCFT.}
\label{factorization}
\end{center}
\end{figure}

A priori, there are various possible resolutions to the different predictions for $G_{\CT_{3,{3\over2}}}$. First, it could be that the extra $SU(2)$ factor is an accidental symmetry at energies $E\ll R^{-1}$ (where $R$ is the radius of the compactification circle). Second, it could be that the 4D description around \eqref{Hfield} from the M5 brane simply misses some flavor symmetries.\footnote{A similar phenomenon occurs in some theories with only regular punctures.} Finally, it could be that neither description gets the correct symmetries.

We claim the 3D quiver of Fig. \ref{quiver3} captures the full flavor symmetry and the 4D description around \eqref{Hfield} does not. In particular, we will argue that the $\CT_{3,{3\over2}}$ SCFT splits into a free hypermultiplet and an interacting theory, $\CT_X$, as in Fig. \ref{factorization} and that the $SU(2)$ symmetry detected around \eqref{Hfield} corresponds to a diagonal subgroup of the $SU(2)^2\subset G_{\CT_{3,{3\over2}}}$ factor. Happily, the interacting $\CT_X$ theory then has ($\CN=2$) flavor symmetry $G_{\CT_X}=SU(3)\times SU(2)$ and the following anomalies\footnote{Somewhat intriguingly, as an $\CN=1$ theory, the flavor symmetry is $SU(3)\times SU(2)\times U(1)$. Note that since the $U(1)$ symmetry comes from the $\CN=2$ $U(1)_R\times SU(2)_R$ symmetry, it is chiral (although the $SU(3)\times SU(2)$ factors are not by the general analysis of \cite{Buican:2013ica}). We are not aware of another method in field or string theory to impose a minimality condition and find $SU(3)\times SU(2)\times U(1)$ as a set of symmetries. However, note that these are genuine (global) symmetries and not gauge symmetries as in the Standard Model.}
\begin{equation}\label{TXanom}
k_{SU(2)}^{\CT_X}=4~, \ \ \ k_{SU(3)}^{\CT_X}=6~, \ \ \ c^{\CT_X}={13\over6}~, \ \ \ a^{\CT_X}={47\over24}~.
\end{equation}
In particular, we can now, in more direct analogy with the $E_6=T_3$ theory, construct conformal manifolds just from arbitrarily many $\CT_X$ theories and conformal gauge fields.\footnote{Since now we can build an infinite linear quiver of $\CT_X$ theories where we alternate gauging $SU(2)$ and $SU(3)$ flavor symmetry factors.} On the other hand, we need to be careful when constructing theories by gauging the $SU(2)$ factor since it has a $\mathbb{Z}_2$ Witten anomaly \cite{Witten:1982fp}! Indeed, as argued in \cite{Buican:2014hfa}, the (diagonal) $\CT_{3,{3\over2}}$ $SU(2)$ factor is anomaly free. However, since a single hypermultiplet has a Witten anomaly, the $\CT_X$ theory must have a non-trivial compensating anomaly.

In order to substantiate our claim in Fig. \ref{factorization} and also to further examine the analogy between the $\CT_X$ theory and the $T_N$ theories, we must go beyond the simple description around \eqref{Hfield}. To that end, we will focus on the \lq\lq Schur" sector \cite{Gadde:2011uv} of the various component theories in our duality. This is a sector of operators that contains a wealth of information and is often exactly solvable, since it contains the (hidden) symmetries of a 2D chiral algebra \cite{Beem:2013sza}.

In order to get a handle on the Schur sector, it is useful to first compute the limit of the superconformal index (i.e., the \lq\lq Schur" index) that captures contributions only from operators in this sector (i.e., the \lq\lq Schur" operators). For our starting point in Fig. \ref{quiver1}, this computation can easily be carried out using the results of \cite{Buican:2015ina,Cordova:2015nma}. Invariance of the Schur index under $S$-duality guarantees that we then also have the index for the theory in Fig. \ref{quiver2}.\footnote{Moreover, the consistency of the resulting picture we will find below bolsters the claimed duality in Fig. \ref{quiver1} and Fig. \ref{quiver2} beyond the checks that were performed in \cite{Buican:2014hfa} at the level of the SW curves and dimensional reductions.}

Obtaining the index of the $\CT_X$ theory itself is somewhat more delicate. However, using a recent conjecture in \cite{Xie:2016evu} (proven in \cite{kac2017remark} and reviewed in Appendix \ref{app:PEproof}), we are able to find the Schur index of $\CT_X$ from the index of the quiver in Fig. \ref{quiver2} using the inversion theorem in \cite{spiridonov2006inversions}. Our use of the result in \cite{spiridonov2006inversions} is in the same spirit that it was used by the authors of \cite{Gadde:2010te} to determine the index of the $E_6$ SCFT (however, there are some technical differences, because our $SU(2)$ duality frame involves an additional strongly interacting factor).

In order to check our index computation and also to gain more insight into the $\CT_X$ theory, we bootstrap its chiral algebra, $\chi(\CT_X)$, (and hence by the correspondence of \cite{Beem:2013sza}, we find its Schur operators) using techniques described in \cite{Lemos:2014lua}. In particular, we show that there is a unique consistent chiral algebra with the (minimal) number of generators required, via the correspondence in \cite{Beem:2013sza}, for compatibility with our inversion result and the anomalies in \eqref{TXanom}. Then, using arguments closely related to those in \cite{Lemos:2014lua}, we argue for an exact expression for the vacuum character of $\chi(\CT_X)$ in terms of certain \lq\lq diagonal" $\widehat{su(2)}_{-2}\times\widehat{su(3)}_{-3}$ Affine Kac-Moody (AKM) characters. By the correspondence of \cite{Beem:2013sza}, this gives us a simple closed-form expression for the Schur index of the $\CT_X$ theory and allows us to recover the $S^3$ partition function of the proposed 3D mirror in Fig. \ref{quiver3} by taking the $q\to1$ limit of this quantity.

As we will see, our expression for the Schur index in terms of AKM characters reveals a much deeper connection with the $T_N$ theories: the \lq\lq structure constants" that emerge are precisely those of the $T_2$ theory (although the AKM characters we sum over are different, they are in one-to-one correspondence with those we sum over in the $T_2$ case). We explore these connections in greater detail below and also comment on some consequences of the non-trivial Witten anomaly of the $\CT_X$ theory for the 2D/4D correspondence of \cite{Beem:2013sza}.

Before proceeding, let us discus the plan of the paper. In the next section, we review the basics of the Schur sector and its correspondence with 2D chiral algebras. With this formalism under our belts, we give a simple argument for the factorization in Fig. \ref{quiver3}. We then move on to describe the Schur index of the $\CT_X$ theory via the $S$-duality of \cite{Buican:2014hfa}. Using this result, we bootstrap the corresponding chiral algebra, construct its vacuum character, and make contact with Fig. \ref{quiver3}. We then compute the Hall-Littlewood index of our theory using the data in Fig. \ref{quiver3} and compare it with our Schur index in order to highlight some subtle aspects of the Schur sector. We conclude with a discussion of various open problems suggested by our work.

\newsec{The Schur sector and the 4D/2D correspondence}\label{2D4Dcorr}
In this section we conduct a lightning review of Schur operators and the parts of the associated 4D/2D correspondence described in \cite{Beem:2013sza} that are useful for us below. These operators sit in short multiplets of the 4D $\CN=2$ superconformal algebra and satisfy
\begin{equation}\label{Schurcond}
\left\{\tilde\CQ_{2\dot-},\CO\right]=\left\{\CQ_{-}^1,\CO\right]=0~,
\end{equation}
along with corresponding equations for the conjugate charges acting on $\CO(0)$. In \eqref{Schurcond}, numerical indices denote spin-half $SU(2)_R\subset U(1)_R\times SU(2)_R$ quantum numbers, while the remaining indices are for spinors of the left and right parts of the Lorentz group. To simplify our notation, we have dropped any $SU(2)_R$ or Lorentz indices of $\CO$, but the above definition guarantees that Schur operators are $SU(2)_R$ and Lorentz highest-weight states satisfying
\begin{equation}
E(\CO)=2R(\CO)+j_1(\CO)+j_2(\CO)~, \ \ \ r(\CO)=j_2(\CO)-j_1(\CO)~,
\end{equation}
where $E$ is the scaling dimension, $R$ is the $SU(2)_R$ weight, $j_{1,2}$ are the Lorentz weights, and $r$ is the $U(1)_R\subset U(1)_R\times SU(2)_R$ charge.

The Schur operators also give the unique contributions to a simpler (but highly non-trivial) limit of the superconformal index called the Schur limit
\begin{equation}
\CI(q, {\bf x})={\rm Tr}_{\CH}(-1)^Fe^{-\beta\Delta}q^{E-R}\prod_i(x_i)^{f_i}~,
\end{equation}
where the trace is over the Hilbert space of local operators, $\CH$, $F$ is fermion number, $|q|<1$ is a superconformal fugacity, the $|x_i|=1$ are flavor fugacities, $f_i$ are flavor charges, and $\Delta=\left\{\tilde\CQ_{2\dot-},\left(\tilde\CQ_{2\dot-}\right)^{\dagger}\right\}$. Schur operators sit in the following multiplets
\begin{equation}\label{multtypes}
\hat\CB_R~, \ \ \ \CD_{R(0,j_2)}\oplus\bar\CD_{R(j_1,0)}~, \ \ \ \hat\CC_{R(j_1,j_2)}~,
\end{equation}
where we have used the notation and conventions of \cite{Dolan:2002zh}.\footnote{See also \cite{Dobrev:1985qv,Minwalla:1997ka}.}

The $\hat\CC_{R(j_1,j_2)}$ multiplets are semi-short multiplets, and the component Schur operators are obtained by acting on the highest-weight state with $\tilde Q_{2\dot+}Q^1_+$. The most important example of such multiplets for us below will be the stress tensor multiplet, $\hat\CC_{0(0,0)}$. The associated Schur operator is the $SU(2)_R$ and Lorentz highest weight component of the $SU(2)_R$ current, $J^{11}_{+\dot+}$.

The $\hat\CB_R$ multiplets will also play an important role below. The corresponding Schur operators are the highest $SU(2)_R$ weight components of the primaries and are annihilated by half the $\CN=2$ superspace. These operators can parameterize the Higgs branch (when it exists). A particularly important example of a $\hat\CB_R$ multiplet is the dimension two $\hat\CB_1$ multiplet. It contains flavor symmetry currents and has as its Schur operator the holomorphic moment map, $\mu$.

The $\CD_{R(0,j_2)}\oplus\bar\CD_{R(j_1,0)}$ multiplets are somewhat less familiar (the component Schur operators are $\tilde Q_{2\dot+}$ and $Q^1_+$ highest-weight descendants),\footnote{Although the case with $R=j_1=j_2=0$ is just the free abelian vector multiplet, and the Schur operators are highest weight gauginos.} but, together with the $\hat\CB_R$ multiplets, the $\bar\CD_{R(j_1,0)}$ multiplets comprise an important subring of operators called the Hall-Littlewood (HL) chiral ring \cite{Gadde:2011uv}. It is an interesting general question to understand the class of theories whose HL ring includes $\CD_{R(0,j_2)}\oplus\bar\CD_{R(j_1,0)}$.\footnote{In the class $\CS$ construction, the existence of these operators can sometimes be related to the topology of the compactification surface, $\CC$ \cite{Gadde:2011uv}.} As we will see below, the HL ring of the $\CT_X$ theory is generated only by operators of type $\hat\CB_R$.

The authors of \cite{Beem:2013sza} found a general organizing principle for all of the above operators: they are related to a 2D chiral algebra. More precisely, the Schur operators define non-trivial cohomology classes with respect to a nilpotent supercharge, $\mathbbmtt{Q}=\CQ^1+\tilde\CS^2$
\begin{equation}
\left\{\mathbbmtt{Q},\CO(0)\right]=0~, \ \ \ \CO(0)\ne\left\{\mathbbmtt{Q},\CO'(0)\right]~.
\end{equation}
One then considers $\CO$ to be fixed in a plane $\mathcal{P}\subset\mathbb{R}^4$ with coordinates $(z, \bar z)$. Translations (and the rest of the global conformal group) in the $\bar z$ direction are twisted with the $SU(2)_R$ symmetries. It then turns out that the quantum numbers of $\CO$ are such that its twisted $\bar z$ translations are $\mathbbmtt{Q}$-exact. In general, this translation process introduces lower $SU(2)_R$ partner components of $\CO$ when $\bar z\ne0$.\footnote{In the notation of \cite{Beem:2013sza}, the twisted-translated Schur operators are written as $\CO(z, \bar z)\equiv u_{i_1}(\bar z)\cdots u_{i_{2N}}(\bar z)\CO^{i_{1}\cdots i_{2N}}$, where $i_k$ are $SU(2)_R$ spin-half indices and $u_{i}\equiv(1, \bar z)$.\label{ttfootnote}} However, these translations do not take one out of the cohomology class defined by $\CO(z,0)$ and so the cohomology classes form an infinite dimensional chiral algebra with meromorphic correlators (translations out of the plane take one out of the cohomology).

While the precise details of the map between 4D and 2D are somewhat technical, the basic results are intuitive. For example, we have the following correspondences \cite{Beem:2013sza}
\begin{equation}\label{map}
\chi\left[J^{11}_{+\dot+}\right]=-{1\over2\pi^2}T~, \ \ \ \chi\left[\mu^I\right]={1\over2\sqrt{2}\pi}J^I~, \ \ \ \chi\left[\partial_{+\dot+}\right]=\partial_z\equiv\partial~,
\end{equation}
where $\chi\left[\cdots\right]$ takes a 4D Schur operator to its 2D counterpart. As one might naturally expect, $T$ is the holomorphic stress tensor, $J^I$ is an AKM current ($I$ is an adjoint index), and $\partial$ is the holomorphic derivative in $\mathcal{P}$. Note that any local 4D $\CN=2$ SCFT has a stress tensor and therefore, by $\CN=2$ SUSY, a $J^{11}_{+\dot+}$ operator. As a result, \eqref{map} tells us that the associated chiral algebra must contain at least a Virasoro sub-algebra. Moreover, 4D theories with flavor symmetries have an associated chiral algebra with an AKM subalgebra. Interestingly, there is a universal map between the corresponding anomalies in 4D and 2D for the universal currents we have just described \cite{Beem:2013sza}
\begin{equation}\label{anomalyMap}
k_{2d}=-{1\over2}k_{4d}~, \ \ \ c_{2d}=-12c_{4d}~.
\end{equation}

More generally, the chiral algebras arising via this correspondence typically contain generators\footnote{Generators are defined to be the operators whose normal-ordered products---along with their derivatives---span the chiral algebra.} beyond the ones appearing in \eqref{map}. However, all generators must satisfy basic consistency conditions in the form of Jacobi identities
\begin{equation}\label{JacobiDef}
\left[\CO_1(z_1)\left[\CO_2(z_2)\CO_3(z_3)\right]\right]-\left[\CO_3(z_3)\left[\CO_1(z_1)\CO_2(z_2)\right]\right]-\left[\CO_2(z_2)\left[\CO_3(z_3)\CO_1(z_1)\right]\right]=0~,
\end{equation}
where we take $|z_2-z_3|<|z_1-z_3|$, $\left[\cdots\right]$ is the singular part of the OPE of the operators enclosed, and we have assumed the $\CO_i$ are all bosonic (as we will see is the case for $\chi(\CT_X)$ below). These constraints are the basis of the chiral algebra bootstrap, and we will make heavy use of them in Sec. \ref{chiboot}.

Finally, we note that the holomorphic dimension in the chiral algebra, $h$, satisfies
\begin{equation}\label{dim}
h=E-R~.
\end{equation}
Moreover, the torus partition function of the chiral algebra can be written as follows
\begin{equation}\label{T2fn}
Z(y,q,{\bf x})={\rm Tr}\ y^{M^{\perp}} q^{L_0}\prod(x_i)^{f_i}~,
\end{equation}
where $M^{\perp}=j_1-j_2$, and the relation to the Schur index is
\begin{equation}\label{SchurT2rel}
Z(-1, q,{\bf x})=\CI(q,{\bf x})~.
\end{equation}
This equation allows us to read off the vacuum character of the chiral algebra from the Schur index and is instrumental in allowing us to find the set of generators of $\chi(\CT_X)$ below.

Therefore, we see that the Schur sector of the theory contains a remarkably constrained---but still interesting---set of operators that are complementary to the Coulomb branch degrees of freedom characterizing the SW curve description discussed in the introduction.\footnote{However, these operators are not independent of the Coulomb branch sector. Indeed, a study of the Schur index of AD theories reveals that the $q\to1$ limit of the index secretly encodes Coulomb branch physics \cite{Buican:2015hsa} (see also related work in \cite{Fredrickson:2017yka}). Moreover, the Schur index can be computed from particular sums over BPS states on the Coulomb branch \cite{Cordova:2015nma}.} Since chiral algebras are such rigid objects, finding a unique chiral algebra with a particular set of generators and anomalies that satisfies Jacobi identities like those in \eqref{JacobiDef} is strong evidence for having found the Schur sector of a 4D theory exactly.

In the next section, we will apply our above discussion and argue for the factorization in Fig. \ref{factorization}. Along the way, we also make use of the results in \cite{Buican:2015ina,Cordova:2015nma}.

\newsec{A chiral algebra argument for $\CT_{3,{3\over2}}=\CT_X\oplus {\rm hyper}$}\label{factarg}
To understand why the $\CT_{3,{3\over2}}$ theory factorizes, note that a simple consequence of the duality discussed in the introduction is that the spectrum of gauge invariant operators arising from the quiver in Fig. \ref{quiver1} must match the spectrum of such operators arising from the quiver in the dual frame in Fig. \ref{quiver2}. In particular, the $SU(3)$ side of the theory clearly has dimension three and $SU(2)_R$ weight ${3\over2}$ baryons
\begin{equation}
B=\epsilon^{ijk}Q_i^aQ_j^bQ_k^c~, \ \ \ \tilde B=\epsilon_{ijk}\tilde Q^i_a\tilde Q^j_b\tilde Q^k_c~,
\end{equation}
that are charged under the baryonic $U(1)\subset U(3)$ factor of the flavor symmetry. Moreover, we have
\begin{equation}
\left[\tilde Q_{2\dot-},B\right]=\left[Q^1_{-},B\right]=\left[\tilde Q_{2\dot-},\tilde B\right]=\left[Q^1_{-},\tilde B\right]=0~,
\end{equation}
and so these degrees of freedom are  Schur operators of type $\hat\CB_{3\over2}$ discussed around \eqref{multtypes}. By \eqref{dim}, Such operators are in turn related to 2D chiral algebra primaries $\CB$ and $\tilde\CB$ of holomorphic scaling dimension $h=E-R={3\over2}$.

As a result, the $SU(2)$ side of the duality must also have operators $B$ and $\tilde B$. Since the $(A_1, D_4)$ factor in this duality frame is responsible for the baryonic symmetry, $\CB$ and $\tilde\CB$ must either be Schur operators of the $(A_1, D_4)$ sector or composite gauge-invariant operators built from Schur operators of this sector and Schur operators of at least one other sector. However, we know the Schur sector of the $(A_1, D_4)$ theory exactly: it corresponds, via the map described in Sec. \ref{2D4Dcorr}, to the $\widehat{su(3)}_{-{3\over2}}$ AKM chiral algebra \cite{Beem:2014zpa,Buican:2015ina,Cordova:2015nma,Buican:2015tda}\footnote{See also the beautiful recent generalization in \cite{Creutzig:2017qyf}.} generated by the AKM current $J^I_{SU(3)}$ ($I=1,\cdots,8$ is an adjoint index of $SU(3)$).

Therefore, $\chi\left[(A_1, D_4)\right]$ has no operators with the quantum numbers of $\CB$ and $\tilde\CB$ (since $J^I_{SU(3)}$ has $h=1$, there are no operators with $h={3\over2}$ in the $\widehat{su(3)}_{-{3\over2}}$ vacuum module). As a result, we must construct $B$ and $\tilde B$ as composites of the holomorphic moment map of the $(A_1, D_4)$ theory, $\mu_{SU(3)}^I$, with a field of dimension one (and $h=1/2$).\footnote{In fact, the baryons map to generators of the chiral algebra related to the theory in Figs. \ref{quiver1} and \ref{quiver2}. Note that, in accord with the bound in \cite{Buican:2016arp},  this chiral algebra has at least three generators, since there are also multiple generators with $h=1$ as well.} In other words, we must have a sector consisting of a hypermultiplet, $Q_i$ (with $i=1,2$), charged under the gauged $SU(2)$ (recall that the hypermultiplet has $Sp(1)\simeq SU(2)$ flavor symmetry) from which we can construct
\begin{equation}
B=\mu_{SU(3)}^iQ_i~,  \ \ \ \tilde B=\tilde\mu_{SU(3)}^iQ_i~,
\end{equation}
where $\mu_{SU(3)}^i$ and $\tilde\mu_{SU(3)}^i$ are the two doublets descending from the eight $\mu_{SU(3)}^I$ moment maps under the decomposition of $SU(3)$ into representations of the $SU(2)$ gauge group (we have ${\bf 8}={\bf1}+2{\bf\times 2}+{\bf3}$). In particular, we see that the $\CT_{3,{3\over2}}$ SCFT splits into a free hyper and another theory which we call $\CT_X$ (as in Fig. \ref{factorization}).\footnote{One may also derive this result using facts about the moduli spaces of vacua for the theories in our duality. However, our arguments at the level of the chiral algebra provide a stronger consistency check of the duality in \cite{Buican:2014hfa} as well as of the picture we propose in Fig. \ref{factorization}.} Moreover, as discussed in the introduction, since the $\CT_{3,{3\over2}}$ theory doesn't have a Witten anomaly for its $SU(2)$ global symmetry subgroup but the free hypermultiplet does, the $SU(2)$ global symmetry subgroup of the $\CT_X$ theory has a Witten anomaly. We will see an interesting consequence of this fact below. This discussion also derives the result in \eqref{TXanom} from \eqref{T332anom}.

In the next section, we begin a deeper exploration of the $\CT_X$ theory. To do so, we first construct the Schur index of the theory. After finding this index, we will conjecture a chiral algebra, $\chi(\CT_X)$, that reproduces it and then use bootstrap techniques to confirm our conjecture.

\newsec{The Schur index of $\CT_X$ from $S$-duality and inversion}\label{inversion}
In order to get more detailed information about the $\CT_X$ theory, we compute its Schur index using the $S$-duality described in Fig. \ref{quiver1} and Fig. \ref{quiver2}. Indeed, since the index is invariant under $S$-duality, the Schur indices of the theories in these two figures must agree. On the $SU(3)$ side of the duality, it is easy to compute the Schur index as follows
\begin{eqnarray}\label{su3side}
\CI_{SU(3)}(q, s, z_1, z_2)&=&\oint d\mu_{SU(3)}(x_1,x_2)\times\CI_{\rm vect}(q,x_1, x_2)\times\CI_{\rm flavors}(q, x_1, x_2, s, z_1, z_2)\times\cr&\times&\CI_{(A_1, D_4)}(q,x_1, x_2)^2~,
\end{eqnarray}
where the measure of integration is the $SU(3)$ Haar measure, $\CI_{\rm flavors}$ is the index of the three fundamental flavors, $\CI_{(A_1, D_4)}$ is the index of the $(A_1, D_4)$ theory, and $\CI_{\rm vect}$ is the vector multiplet index (see Appendix \ref{inversionap} for detailed expressions). The fugacities, $s$ and $(z_1,z_2)$, are for $U(1)\subset U(3)$ and $SU(3)\subset U(3)$ flavor subgroups, respectively. All terms appearing in the integrand of \eqref{su3side} have known closed-form expressions ($\CI_{(A_1, D_4)}$ was computed in \cite{Buican:2015ina,Cordova:2015nma}). Now, on the $SU(2)$ side of the duality, we have
\begin{equation}\label{su2side}
\CI_{SU(2)}(q,s,z_1,z_2)=\oint d\mu_{SU(2)}(e)\times\CI_{\rm vect}(q,e)\times\CI_{(A_1, D_4)}(q,e,s)\times \CI_{\CT_{3,{3\over2}}}(q,e,z_1,z_2)~,
\end{equation}
where $\CI_{\CT_{3,{3\over2}}}$ is the Schur index of the $\CT_{3,{3\over2}}$ theory. From the general discussion in the previous section and Fig. \ref{factorization}, we must have
\begin{equation}\label{Ifact}
\CI_{\CT_{3,{3\over2}}}(q,e,z_1,z_2)=\CI_{\CT_{X}}(q,e,z_1,z_2)\times \CI_{\rm hyper}(q,e)~,
\end{equation}
where the second factor on the RHS is the Schur index of a free hypermultiplet, and the first factor is the index of the $\CT_X$ SCFT.

In order to compute the index in \eqref{Ifact}, we will use an inversion procedure based on the theorem in \cite{spiridonov2006inversions} to extract it from the expression in \eqref{su2side}. Roughly the same basic procedure was first used in \cite{Gadde:2010te} to extract the index of the $E_6$ SCFT  from Argyres-Seiberg duality. However, there are some technical differences (due to the fact that our $SU(2)$ duality frame has an additional strongly interacting factor) in our use of \cite{spiridonov2006inversions} that are reviewed in Appendix \ref{inversionap}. One important precondition for our inversion procedure involves the use of a conjectured form for $\CI_{(A_1, D_4)}(q,x_1, x_2)$ due to Xie-Yan-Yau (XYY) \cite{Xie:2016evu} (recently proved in \cite{kac2017remark} and reviewed in Appendix \ref{app:PEproof}) that is compatible with its known form in \cite{Buican:2015ina,Cordova:2015nma}
\begin{equation}\label{XYYform}
\CI_{(A_1, D_4)}(q,x_1,x_2)=P.E.\left[{q\over1-q^2}\chi_{\rm Adj}(x_1, x_2)\right]~,
\end{equation}
where the \lq\lq plethystic exponential" is defined as
\begin{equation}\label{PEdef}
P.E.\left[G(a_1,\cdots,a_p)\right]\equiv\exp\left[\sum_{n=1}^{\infty}{1\over n}G(a_1^n,\cdots, a_p^n)\right]~,
\end{equation}
for any function of the fugacities, $G$. Indeed, the surprising fact that the index of the strongly interacting $(A_1, D_4)$ SCFT in \eqref{XYYform} is related to the index of a free adjoint hypermultiplet by the rescaling $q\to \sqrt{q}$ allows us to use the inversion theorem of \cite{spiridonov2006inversions} (as in \cite{Gadde:2010te}, we will a posteriori justify the assumptions used in applying this theorem by finding a consistent symmetry structure for our index). One surprising fact we will uncover later on is that, when appropriately re-written, $\CI_{\CT_X}$ will also be closely related to a Schur index for free fields.

Applying the procedure in Appendix \ref{inversionap}, we find that the Schur index of the $\CT_{3,{3\over2}}$ theory can be written as
\begin{equation}\label{schurtotal}
\CI_{\CT_{3,{3\over2}}}(q,w,z_1,z_2)=\frac{1}{(w^{\pm 2} q;q)}\left[ \frac{1}{1-w^2}\CI_{SU(3)}(q,wq,z_1,z_2)+\frac{w^2}{w^2-1}\CI_{SU(3)}(q,\frac{q}{w},z_1,z_2)\right]~,
\end{equation}
where $(a;q)$ denotes the $q$-Pochhammer symbol
\begin{equation}
(a;q)=\prod_{n=0}^{\infty} (1-a q^n)~,
\end{equation}
and we also use the condensed notation
\begin{equation}\label{repeat}
(a^{\pm};q)\equiv (a;q)(a^{-1};q)~.
\end{equation}
Expanding \eqref{schurtotal} perturbatively in $q$ we obtain
\begin{eqnarray}\label{schurT332}
\CI_{\CT_{3,{3\over2}}}(q, w, z_1, z_2)&=&1+\chi_{1}q^{1\over2}+(2 \chi_{2} + \chi_{1,1})q+2(\chi_{1} + \chi_{3} + \chi_{1} \chi_{1,1})q^{3\over2}+(4 + 3 \chi_{2} +\cr&+& 3 \chi_{4} + 3 \chi_{1,1} + 
 3 \chi_{2} \chi_{1,1} + \chi_{2,2})q^2+(8 \chi_{1} + 5 \chi_{3} + 3 \chi_{5} +  7 \chi_{1} \chi_{1,1} +\cr&+& 
 4 \chi_{3} \chi_{1,1} + \chi_{1} \chi_{3, 0} + \chi_{1} \chi_{0, 3} + 2 \chi_{1} \chi_{2,2})q^{5\over2}+(6 + 15 \chi_{2} + 6 \chi_{4} +\cr&+& 4 \chi_{6} + 10 \chi_{1,1} +  12 \chi_{2} \chi_{1,1} + 5 \chi_{4} \chi_{1,1} +  3 \chi_{3, 0} + \chi_{2} \chi_{3, 0} + 3 \chi_{0,3} +\cr&+& \chi_{2} \chi_{0, 3} + 3 \chi_{2,2} +  4 \chi_{2} \chi_{2,2} + \chi_{3,3})q^3+\CO(q^{7\over2})~,
\end{eqnarray}
where  $\chi_{\lambda}\equiv\chi_{\lambda}(w)$ is the character of the spin ${\lambda\over2}$ representation of $SU(2)$ and $\chi_{\lambda_1,\lambda_2}\equiv\chi_{\lambda_1,\lambda_2}(z_1, z_2)$ is the character of the $SU(3)$ representation with Dynkin labels $\lambda_{1,2}\in\mathbb{Z}_{\ge0}$. 

One check of \eqref{schurtotal} and \eqref{schurT332} is that they are compatible with the factorization we argued for in Sec. \ref{factarg} and explained at the level of the index in \eqref{Ifact}. In particular, we see a free hypermultiplet at $\CO(q^{1\over2})$. Moreover, the total global symmetry of the $\CT_{3,{3\over2}}$ theory is then, as explained in the introduction, $SU(2)^2\times SU(3)$ with one $SU(2)$ factor coming from the free hypermultiplet.\footnote{Note that, on the $SU(2)$ side of our duality, we gauge the diagonal $SU(2)\subset SU(2)^2$ to construct the theory in Fig. \ref{quiver2}.} Although this enhancement is not quite as dramatic as the $E_6$ enhancement of flavor symmetry observed in the example studied in \cite{Gadde:2010te}, we will find a much deeper statement about the (hidden) symmetries of this theory (and hence the consistency of our picture) by bootstrapping the chiral algebra associated with $\CT_X$ below.

As a first step towards this goal, we arrive at the index of the $\CT_X$ theory by dividing both sides of \eqref{Ifact} by the free hypermultiplet contribution
\begin{eqnarray}\label{schurTX}
\CI_{\CT_X}(q,w,z_1,z_2)&=&1+(\chi_{2}+\chi_{1,1})q+\chi_{1} \chi_{1,1} q^{3\over2}+(2 + \chi_{2} + \chi_{4} +  2 \chi_{1,1} + \chi_{2} \chi_{1,1} + \cr&+&\chi_{2,2})q^2+(\chi_{1} + 2 \chi_{1} \chi_{1,1}+ \chi_{3} \chi_{1,1} + \chi_{1} \chi_{0,3} + \chi_{1} \chi_{3,0} +\cr&+& \chi_{1} \chi_{2,2})q^{5\over2}+(2 + 4 \chi_{2} + \chi_{4} + \chi_{6} + 5 \chi_{1,1} +3 \chi_{2} \chi_{1,1} + \chi_{4} \chi_{1,1} +\cr&+& 2 \chi_{3, 0} + 2 \chi_{0,3} + 2 \chi_{2,2} + 2 \chi_{2} \chi_{2,2} + \chi_{3,3})q^3+\CO(q^{7\over2})~,
\end{eqnarray}
which has, as promised, $SU(2)\times SU(3)$ global symmetry (we see currents in the adjoint of this symmetry group at $O(q)$, and the index organizes into characters of this symmetry). 

In the next section, we use this expansion to conjecture the generators of the associated chiral algebra, $\chi(\CT_X)$. We then bootstrap this chiral algebra and show that it is consistent (in the sense that it obeys Jacobi identities of the form reviewed in \eqref{JacobiDef}). Moreover, we will argue that it is the unique such chiral algebra with the generators we conjecture and the anomalies required from the discussion in the introduction and Sec. \ref{2D4Dcorr}.\footnote{We will also see that, for example, the central charge of the chiral algebra is fixed to be $c_{2d}=-26$ given our generators and AKM levels. Similarly, the AKM levels are fixed given our generators and $c_{2d}=-26$ (here we assume that the 2D chiral algebra is related to a unitary 4D SCFT by the correspondence of \cite{Beem:2013sza}).}

\newsec{A chiral algebra conjecture}
From the simple expansion presented in \eqref{schurTX}, we can immediately conjecture the generators of the corresponding chiral algebra in the sense of \cite{Beem:2013sza} reviewed in Sec. \ref{2D4Dcorr}. Indeed, using the map in \eqref{SchurT2rel}, \eqref{schurTX} is also an expansion for the character of the vacuum module of the chiral algebra we want to find.

The only possible contributions in the vacuum module at $\CO(q)$ must come from AKM currents, which, in this case, are for $\widehat{su(2)}_{-2}\times\widehat{su(3)}_{-3}$. We have used \eqref{TXanom} and \eqref{anomalyMap} to fix the levels of the AKM algebras to the so-called critical levels (these are $k=-h^{\vee}$, where $h^{\vee}$ is the dual Coxeter number). As in the case of the $T_N$ theories (with the exception of the $T_3=E_6$ theory which has enhanced $E_6\supset SU(3)^3$ flavor symmetry and the $T_2$ theory which has $Sp(4)\supset SU(2)^3$ symmetry and no AKM currents as generators), this discussion means that the holomorphic stress tensor of the chiral algebra must be an independent generator, since the Sugawara stress tensor is not normalizable (note that from \eqref{TXanom} and \eqref{anomalyMap} we have $c=-26$ for the Virasoro subalgebra\footnote{Amusingly, this value is the same as the $c$ anomaly for the $bc$ ghost system.}). Looking at $\CO(q^{3\over2})$, we see that there must be at least one operator, $\CO_{aI}$, transforming in the ${\bf2}\times{\bf8}$ representation of the global symmetry (since all the other generators are integer dimensional).\footnote{This operator must be of type $\hat\CB_{3\over2}$. The only other Schur multiplets (see \eqref{multtypes}) of the appropriate statistics that can appear at $\CO(q^{3\over2})$ are $\CD_{0(0,{1\over2})}\oplus\bar\CD_{0({1\over2},0)}$. However, these operators have the wrong multiplicity and, on general grounds, should not be present in this theory \cite{Buican:2014qla} (note that they also satisfy free field equations of motion and so presumably should not appear on such grounds as well).} This operator is mapped to an AKM primary, $\chi[\CO_{aI}]=W_{aI}$. Therefore, the minimal conjecture for $\chi\left[\CT_X\right]$ is the following

\medskip
\label{conjecture}
\noindent
{\bf Conjecture:} The chiral algebra, $\chi\left[\CT_X\right]$, is generated by a stress tensor, $T$ (with $c=-26$), AKM currents, $J_{su(2)}^A$ and $J_{su(3)}^I$ (with $A=1,\cdots,3$ and $I=1,\cdots,8$) for $\widehat{su(2)}_{-2}\times\widehat{su(3)}_{-3}$, and an $h={3\over2}$ AKM primary, $W_{aI}$ (with $a=1,2$ and $I=1,\cdots,8$), transforming in the ${\bf 2}\times{\bf8}$ representation of $su(2)\times su(3)$.
\medskip

Note that this conjecture is consistent with the simplicity of AD theories: to get the chiral algebra of $\CT_X$, one needs to add only a single additional generator (really 16 generators if one counts all the allowed $a,I$ pairs) beyond the universal ones required by 4D symmetries. Indeed, this algebra is considerably simpler than those of the interacting $T_N$ theories (even the $T_3=E_6$ theory has a larger number of generators by virtue of its large global symmetry). 

We will give convincing evidence for this conjecture in Sec. \ref{chiboot}, where we will show there is a unique consistent chiral algebra satisfying this conjecture. For now, we also give some powerful circumstantial evidence in favor of our proposal. In particular, if this conjecture is correct, then all contributions appearing in \eqref{schurTX} can be generated by plethystic exponentials of our generators modulo constraints. Assuming our conjecture is correct, we find some natural operator relations at low order in $q$

\begin{itemize}
\item{A singlet relation at $O(q^2)$. As we will see in greater detail below, we expect that
\begin{equation}\label{flavornull2}
{\rm Tr}J_{SU(3)}^2\sim{\rm Tr}J_{SU(2)}^2~,
\end{equation}
where we will fix the non-zero constant of proportionality in the next section. The motivation for this relation is that the $\widehat{su(2)}_{-2}\times\widehat{su(3)}_{-3}$ subalgebras of $\chi(\CT_X)$ are at the critical level. Therefore, in their respective modules, the LHS and RHS of \eqref{flavornull2} separately vanish. However, it is natural to expect that, as in the case of the $T_N$ theories \cite{Lemos:2014lua}, one linear combination of these operators becomes non-null in the full chiral algebra and therefore remains as a non-trivial operator.}
\item{At $\CO(q^{5\over2})$ we have two operator relations with quantum numbers ${\bf 2}\times {\bf 8}$}.
\item{At $\CO(q^3)$ we have many operator relations. One important set of relations are the singlets of the form
\begin{equation}\label{t3sing}
{\rm Tr}J_{SU(3)}^3={\rm Tr}J_{SU(2)}^3=W_{3\over2}^{aI}W_{{3\over2} aI}=0~.
\end{equation}
The first relation again follows from the fact that the flavor symmetry is at the critical level and is a non-trivial statement, while the last two relations are a simple consequence of bosonic statistics.}
\end{itemize}

\newsec{Bootstrapping the Chiral Algebra of $\CT_X$}\label{chiboot}
One strong piece of evidence in favor of our conjecture in the previous section is that there exists a (unique) set of operator product expansions (OPEs) among the generators described there that is consistent with Jacobi identities of the type described in \eqref{JacobiDef}. To understand this statement, let us consider the most general OPEs among the generators. The non-vanishing singular parts of the OPEs among the stress tensor and the $SU(2)\times SU(3)$ currents are completely fixed by Ward identities to take the form
\begin{align}
T(z)T(0) &\sim \frac{c_{2d}}{2z^4} + \frac{2T}{z^2} + \frac{\partial T}{z}~,\nonumber
\\
T(z)J^A_{SU(2)}(0) &\sim \frac{J^A_{SU(2)}}{z^2} + \frac{\partial J^A_{SU(2)}}{z}~,\nonumber
\\
T(z) J^I_{SU(3)}(0) &\sim \frac{J^I_{SU(3)}}{z^2} + \frac{\partial J^I_{SU(3)}}{z}~,
\\
J_{SU(2)}^A(z) J_{SU(2)}^B(0) &\sim \frac{k_{2d}^{su(2)}\delta^{AB}}{2z^2} + \frac{i\epsilon^{ABC}J_{SU(2)}^C}{z}~,\nonumber
\\
J_{SU(3)}^I(z)J_{SU(3)}^J &\sim \frac{k_{2d}^{su(3)}\delta^{IJ}}{2z^2} + \frac{if^{IJK}J_{SU(3)}^K}{z}~,\nonumber
\end{align}
where $f_{IJK}$ is the structure constant of $su(3)$ and, as discussed in the previous section, $c_{2d}=-26,\, k_{2d}^{su(2)} = -2$ and $k_{2d}^{su(3)}=-3$. Moreover, since there is no generator with $h=1/2$, $W_a{}^I$ has to be a primary of the Virasoro and $\widehat{su(2)}_{-2}\times \widehat{su(3)}_{-3}$ algebras. This fact implies the following singular parts of the OPEs:
\begin{align}
T(z)W_a{}^I(0) &\sim \frac{3W_a{}^I}{2z^2} + \frac{\partial W_a{}^I}{z}~,\nonumber
\\
J^A_{SU(2)}(z)W_a{}^I(0) &\sim \frac{ \sigma^A_{ab} W^{bI}}{2z}~,
\\
J^I_{SU(3)}(z)W_a{}^J(0) &\sim \frac{f^{IJK}W_{aK}}{z}~,\nonumber
\end{align}
where the $\sigma^A$ are Pauli matrices.

On the other hand, the OPE between $W_a{}^I$ and $W_b{}^J$ is not fixed by the symmetries. Therefore, we adopt the following general ansatz for the singular parts of this OPE:
\begin{align}
W_a{}^I(z) W_b{}^J(0) &\sim \frac{\epsilon_{ab}\delta^{IJ}}{z^3} + \frac{1}{z^2}\bigg(\frac{a_1}{2} \delta^{IJ} \sigma^A_{ab}\, J_{SU(2)A} + \epsilon_{ab}\,(a_2\, f^{IJK} + a_3\, d^{IJK}) J_{SU(3)K}\bigg) 
\nonumber\\
& + \frac{1}{z}\Bigg[ \epsilon_{ab}\delta^{IJ}\bigg(a_4\, T+a_5\, J^A_{SU(2)}J_{SU(2)A} + a_6\, J^K_{SU(3)}J_{SU(3)K} \bigg) 
\nonumber\\
&\qquad + \frac{a_7}{2}\,\delta^{IJ}\sigma^A_{ab}\,J_{SU(2)A}' + a_8\, \epsilon_{ab}\,f^{IJK}\, J_{SU(3)K}'+\frac{a_9}{2} \,\sigma^A_{ab}\,f^{IJK}J_{SU(2)A} J_{SU(3)K}
\nonumber\\
&\qquad + \epsilon_{ab}(a_{10}\, f^{IJK}+a_{11}\,d^{IJK})d_{KLM}J^{L}_{SU(3)}J^M_{SU(3)} + 2a_{12}\,\epsilon_{ab}J^{(I}_{SU(3)}J^{J)}_{SU(3)}\Bigg]~,
\label{eq:WW-OPE}
\end{align}
where $d^{IJK}$ is the totally symmetric tensor of $su(3)$ normalized so that $d^{IJK}d_{IJK}=\frac{40}{3}$, and the $W_a{}^I$ are normalized so that the coefficient of $\epsilon_{ab}\delta^{IJ}/z^3$ is one.\footnote{Note that the coefficient of $\epsilon_{ab}\delta^{IJ}/z^3$ is non-vanishing because otherwise $W_a{}^I$ is null. Therefore, this normalization is always possible.} The twelve coefficients, $a_1,\cdots,a_{12}$, are free parameters to be fixed in such a way that the Jacobi identities are satisfied.
Note that \eqref{eq:WW-OPE} is the most general OPE written in terms of the generators, $T,\, J^A_{SU(2)},\, J^I_{SU(3)}$ and $W_a{}^I$.\footnote{In particular, note that $J^{[I}_{SU(3)}J^{J]}_{SU(3)}$ is vanishing and therefore does not appear as an independent term.}

To fix the above constants and test the consistency of what we have written, we impose the various Jacobi identities among the generators. In particular, the Jacobi identities among $\mathcal{O},\,W_a{}^I$, and $W_b{}^J$ for $\mathcal{O}\in\left\{T,\, J_{SU(2)}^A,\,J_{SU(3)}^I\right\}$ imply that
\begin{align}
a_1 = 1~,\quad a_2 =a_9&= -\frac{2i}{3}~,\quad a_3=a_{10}=0,\quad a_4=-\frac{1}{4}~,\quad a_6=\frac{2-3a_5}{12}~,
\nonumber\\[2mm]
&  a_7=a_{11}=\frac{1}{2}~,\quad a_8=-\frac{i}{3}~, \quad  a_{12}=-\frac{1}{12}~.
\label{eq:solution}
\end{align}
Note that this condition fixes all the OPE coefficients except for $a_5$. Moreover, it turns out that, with $a_6=(2-3a_5)/12$ imposed, the undetermined parameter $a_5$ is only coupled to a null operator. Indeed, under the condition $a_6=(2-3a_5)/12$, the only $a_5$-dependent term in \eqref{eq:WW-OPE} is
\begin{align}
a_5 \left(J_{SU(2)}^A J_{SU(2)A} - \frac{1}{4}J_{SU(3)}^K J_{SU(3)K}\right)~.
\label{eq:null1}
\end{align}
Since the OPEs of this operator with the generators only involve operators of holomorphic dimension larger than or equal to its own dimension, \eqref{eq:null1} is a null operator. Therefore, we set $a_5=0$ in the rest of this section. 

Let us now look at the Jacobi identities among $W_a{}^I,\,W_b{}^J,$ and $W_c{}^K$. With the condition \eqref{eq:solution}, they are automatically satisfied up to the following operators:
\begin{align}
\sigma^A_{ab}J_{SU(2)A}W^{bI} + \frac{if^{IJK}}{2}J_{SU(3)J}W_{aK}~,\ \ \ d^{IJK}J_{SU(3)J}W_{aK}~.
\label{eq:null2}
\end{align}
Since the OPEs of these operators with the generators of the chiral algebra only involve operators of holomorphic dimensions larger than or equal to their own dimensions, the above two operators are both null. This means that \eqref{eq:solution} is consistent with all the Jacobi identities among the generators. The existence of such a consistent $WW$ OPE is strong evidence for our chiral algebra conjecture in the previous section.

Another interesting observation is that the chiral algebra generated by $T,\,W_a{}^I$, and $J_{SU(2)}^A,\, J_{SU(3)}^I$ at the critical levels exist if and only if the Virasoro central charge is $c_{2d}=-26$. Indeed, when we do the above analysis with $c_{2d}$ unfixed, we see that the Jacobi identities among the generators imply $c_{2d}=-26$. Similarly, if we take $c_{2d}=-26$ with the levels of the AKM algebras unfixed, we can show that the Jacobi identities imply that $k_{SU(2)}=-2$ and $k_{SU(3)}=-3$.\footnote{This last statement is true as long as the 2D chiral algebra is related to a unitary 4D SCFT by the correspondence discussed in \cite{Beem:2013sza}.}

We have seen there are at least three null operators up to $h=\frac{5}{2}$. The first one is shown in \eqref{eq:null1} and is a singlet of $SU(2)\times SU(3)$ with $h=2$. The second and third null operators are shown in \eqref{eq:null2} and are in the ${\bf 2}\times {\bf 8}$ representation of $SU(2)\times SU(3)$ with $h=\frac{5}{2}$. These three null operators are perfectly consistent with the 4D operator relations discussed in Sec.~\ref{conjecture}.

Finally, we note that the following normal-ordered product
\begin{equation}\label{JWnon0}
J^I_{SU(3)}W^a_I\ne0~,
\end{equation}
does not vanish. On the other hand, as we will see below when we discuss the HL chiral ring, there is a non-trivial operator relation for the 4D $\hat\CB_R$ ancestors of these operators. However, as we will explain in greater detail below, this statement is consistent with \eqref{JWnon0} because of the $SU(2)_R$ mixing described in Footnote \ref{ttfootnote} which induces a non-trivial $\hat\CC_{{1\over2}(0,0)}$ component for the chiral algebra normal-ordered product.\footnote{Therefore, the Schur operator sitting in this $\hat\CC_{{1\over2}(0,0)}$ multiplet does not map to a generator of the chiral algebra. This situation is quite similar to what happens in, say, the chiral algebra of the $T_3=E_6$ theory, where the stress tensor is not a new generator of $\chi(E_6)$ due to the $SU(2)_R$ twisting of the moment maps and the mixing in of the $\hat\CC_{0(0,0)}$ multiplet in the corresponding normal-ordered product.}

Given this chiral algebra, we will argue that its vacuum character has a surprisingly simple exact expression in terms of certain $\widehat{su(2)}_{-2}\times\widehat{su(3)}_{-3}$ characters. This expansion will turn out to be remarkably similar to the expansion one finds for the $T_2$ theory (although the precise characters we sum over are different). In addition to pointing to some mysterious connections between AD theories and $T_N$ SCFTs, we are able to use this formula to take the $q\to1$ limit and make contact with the $S^3$ partition function of the 3D quiver appearing in Fig. \ref{quiver3}.

\newsec{Re-writing the index in terms of AKM characters}
Since $\chi\left[\CT_X\right]$ has AKM symmetry, it is reasonable to organize the index in terms of AKM representations. In particular, we claim that \eqref{schurTX} can be re-written as follows
\begin{align} \label{2d partition}
I_{\mathrm{\CT_X}}(q,w,z_1,z_2)=\sum_{\lambda=0}^\infty q^{\frac{3}{2} \lambda} P.E.\left[\frac{2q^2}{1-q}+2q-2q^{\lambda+1}\right]\mathrm{ch}^{SU(2)}_{R_\lambda}(q,w)\mathrm{ch}^{SU(3)}_{R_{\lambda,\lambda}}(q,z_1,z_2)~,	
\end{align}
where ${\rm ch}_{R_{\lambda}}^{SU(2)}$ and ${\rm ch}_{R_{\lambda,\lambda}}^{SU(3)}$ are AKM characters with highest-weight states transforming in representations of $SU(2)$ and $SU(3)$ characterized by Dynkin labels $\lambda$ and $\lambda_1=\lambda_2=\lambda$ respectively.

In fact, \eqref{2d partition} is a completely explicit formula, since AKM characters of $\widehat{su}(N)$ at the critical level have the following simple closed-form expression (e.g., see \cite{Lemos:2014lua})
\begin{align}
\mathrm{ch}_{R_{\vec{\lambda}}}(\boldsymbol{x})=\frac{\mathrm{P.E.}[\frac{q}{1-q}\chi_{adj}(\boldsymbol{x})]\chi_{R_{\vec{\lambda}}}(\boldsymbol{x})}{q^{\braket{\vec{\lambda},\rho}}\mathrm{P.E.}[\sum_{j=1}^{N-1}\frac{q^{j+1}}{1-q}]\dim_q R_{\vec{\lambda}}}~,
\end{align} 
where $\vec{\lambda}$ is a vector containing the $N-1$ Dynkin labels characterizing the $su(N)$ quantum numbers of the highest-weight state, $\rho$ is the Weyl vector, $\langle\cdot,\cdot\rangle$ is the standard inner product,\footnote{For $SU(N)$, we have $\langle\vec{\lambda},\rho\rangle=\sum_{i,j}\lambda_iF^{ij}\rho_j=\sum_{i,j}\lambda_iF^{ij}$ (where we have used that $\rho=(1,\cdots,1)$ in the last step) and $F^{ij}$ is the quadratic form matrix (i.e., the inverse of the Cartan matrix). In the cases of interest, this inner product reduces to
\begin{equation} 
\braket{\lambda,\rho}_{SU(2)}=\frac{1}{2}\lambda_1~,\ \ \  \braket{\vec{\lambda},\rho}_{SU(3)}=\lambda_1+\lambda_2~.
\end{equation}} and the $q$-dimension is defined as
\begin{equation}
\dim_q R_{\vec{\lambda}}=\prod_{\alpha \in \Delta_+}\frac{\left[\braket{\vec{\lambda}+\rho,\alpha} \right]_q}{\left[\braket{\rho,\alpha} \right]_q}~,
\end{equation}
where $\Delta_+$ denotes the set of positive roots, and the $q$-deformed number is given by 
\begin{equation}
[x]_q=\frac{q^{-\frac{x}{2}}-q^{\frac{x}{2}}}{q^{-\frac{1}{2}}-q^{\frac{1}{2}}}~.
\end{equation}

Amusingly, we can give an argument in favor of \eqref{2d partition} that parallels the discussion in \cite{Lemos:2014lua} for the $T_N$ case. The first term, $q^{{3\over2}\lambda}$, is related to the dimension of the non-trivial AKM primary, $W^a_I$,  and the dimensions of its products. The plethystic exponential \lq\lq structure constants"
\begin{equation}\label{PEakm}
P.E.\left[{2q^2\over1-q}+2q-2q^{\lambda+1}\right]~,
\end{equation}
have a simple interpretation as well. Indeed, the first term adds in normal-ordered products of the stress tensor and its derivatives with the other operators in the theory (note that these operators vanish in the AKM modules at the critical level) and also adds in normal-ordered products of the $h=2$ state built out of Casimirs of currents orthogonal to \eqref{eq:null1} with other operators in the theory (since this linear combination should not be null in the full chiral algebra). The second term in \eqref{PEakm} adds back in the level one modes of these two operators, and the final term subtracts relations (for $\lambda=0$, this relation is required by the invariance of the vacuum under these modes).

We have also conducted many highly non-trivial checks of \eqref{2d partition}. For example, we have checked that, perturbatively in $q$, \eqref{2d partition} coincides with the expression in \eqref{schurTX} to very high order. Non-perturbatively in $q=e^{-\beta}$ we have also performed various checks. For example, it is straightforward to see that
\begin{equation}\label{cardyTX}
\lim_{\beta\to0}\, \log\, \CI_{\CT_X}(q, w, z_1, z_2)={5\pi^2\over3\beta}+\cdots~.
\end{equation}
This behavior is consistent with the expected Cardy-like scaling discussed in \cite{DiPietro:2014bca,Ardehali:2015bla,DiPietro:2016ond}\footnote{Such behavior holds for theories whose $S^3$ partition function (upon performing an $S^1$ reduction) is finite. On the other hand, we are not aware of any $\CN=2$ SCFT counterexamples to this behavior. Moreover, this scaling has been observed in many classes of strongly interacting $\CN=2$ SCFTs \cite{Buican:2015ina,Buican:2017uka}.}
\begin{equation}
\lim_{\beta\to0}\,\log\, \CI(q,{\bf x})=-{8\pi^2\over3\beta}(a-c)+\cdots={\pi^2\over3\beta}{\rm dim}_{\mathbb{Q}}\CM_H+\cdots~,
\end{equation}
where, the last equality holds by $U(1)_R$ 't Hooft anomaly matching in theories with genuine Higgs branches (i.e., moduli spaces where, at generic points, the theory just has free hypermultiplets). In the case of the $\CT_X$ theory, we expect there to be a genuine Higgs branch since the mirror of the $S^1$ reduction of the $\CT_{3,{3\over2}}$ theory in Fig. \ref{quiver3} has a genuinie Coulomb branch (the result in \eqref{cardyTX} can also be taken as further evidence for the proposal in Fig. \ref{quiver3}).

An even more interesting non-perturbative in $q$ check of our above discussion is to take the $\beta\to0$ limit of \eqref{2d partition}, drop the divergent piece in \eqref{cardyTX}, and study the resulting $S^3$ partition function, $Z_{S^3}$. As we review in greater detail in Appendix \ref{S3partfn}, using the prescription in \cite{Nishioka:2011dq} we obtain
\begin{eqnarray}\label{beta0TX}
\lim_{\beta\to0}\CI_{\CT_X}(q, w, z_1, z_2)&=&{\rm Div.}\times\int_{-\infty}^{\infty} \frac{d m}{\sinh 2 \pi m \sinh \pi m}{\sin\pi m(\zeta_1-\zeta_2)\sin\pi m(2\zeta_1+\zeta_2)\over\sinh\pi(\zeta_1-\zeta_2)\sinh\pi(2\zeta_1+\zeta_2)}\cr&\times&{\sin\pi m(2\zeta_2+\zeta_1)\sin2\pi m\zeta\over\sinh\pi(2\zeta_2+\zeta_1)\sinh2\pi\zeta}~,
\end{eqnarray}
where the \lq\lq Div." factor is the flavor-independent divergent piece in \eqref{cardyTX}, $w=e^{-i\beta\zeta}, \ z_k=e^{-i\beta\zeta_k}$, and the summation over $\lambda$ in \eqref{2d partition} becomes an integral over $m$. On the other hand, we can compute the partition function of the mirror of the quiver in Fig. \ref{quiver3}, given in Fig. \ref{quiver5} of Appendix \ref{S3partfn}, (or of the original quiver in Fig. \ref{quiver3} itself) and divide out by the contribution of a decoupled hypermultiplet to obtain
\begin{eqnarray}
Z_{S^3}^{\rm quiver}&=&{\rm Div.}\times{1\over2}\int_{-\infty}^{\infty} dx_1 dx_2 \frac{\sinh^2(\pi(x_1 - x_2)) e^{2\pi i \eta(x_1 + x_2)}}{\cosh\pi(x_1 - x_2 - m’) \cosh\pi(x_2 - x_1 - m’)} \cr&\times&\frac{1}{\cosh\pi m'\cosh\pi(x_1 - m_1)  \cosh\pi(x_2 - m_1) \cosh\pi(x_1 - m_2)}\cr&\times&\frac{1}{\cosh\pi(x_2 - m_2) \cosh\pi(x_1 + m_1 + m_2) \cosh\pi(x_2 + m_1 + m_2)}~.
\end{eqnarray}
A direct calculation carried out in further detail in Appendix \ref{S3partfn} reveals that (up to an unimportant overall constant)
\begin{equation}
\lim_{\beta\to0}\left({\rm Div.}^{-1}\times \CI_{\CT_X}\right)=Z_{S^3}^{\rm quiver}~,
\end{equation}
when we identify $m_i\leftrightarrow\zeta_i$ and $m'\leftrightarrow\zeta$.\footnote{The fact that there are no imaginary FI parameters turned on is consistent with the 4D $U(1)_R$ symmetry flowing to the Cartan of the 3D $SU(2)_L\subset SO(4)_R$. This statement is also consistent (at least as far as the $\CN=2$ chiral operators of the $\CT_X$ theory are concerned) with the $SU(2)$ quantization condition discussed in \cite{Buican:2015hsa}.} This result is a strong check of our discussion and also of the proposal in \cite{Xie:2012hs,Xie:2013jc}.

In the next section we move on and discuss the HL limit of the index and some additional predictions for the Schur sector of $\CT_X$. Before doing so, let us make a few brief comments on what we have found in this section
\begin{itemize}
\item{The structure constants given in \eqref{PEakm} that multiply the AKM characters in \eqref{2d partition} are precisely those of the free $T_2$ theory \cite{Lemos:2014lua}. While the set of modules we sum over is \lq\lq diagonal," it is not the same set of modules we sum over for the $T_2$ theory (although the modules are in one-to-one correspondence). It is quite remarkable that all the component Schur indices in our duality described in Fig. \ref{quiver1} and \ref{quiver2} are so closely related to those of free fields. Moreover, the form of the partition function in \eqref{2d partition} suggests simple generalizations to other (hypothetical) SCFTs.}
\item{We have found strong evidence in favor of the quiver given in Fig. \ref{quiver3} for the mirror of the $S^1$ reduction of the $\CT_{3,{3\over2}}$ theory. Note, however, that the corresponding mirror for the $S^1$ reduction of the $\CT_X$ theory contains 3D monopole mass terms\footnote{We thank S.~Benvenuti and S.~Giacomelli for a discussion on this point.}
\begin{equation}\label{N2massdef}
\delta W_{\CN=2}=m\varphi_+\CO_++m\varphi_-\CO_-~,
\end{equation}
where $\CO_{\pm}$ are the monopoles in the UV theory that map to the free (twisted) hypermultiplet according to the discussion in Footnote \ref{topology}, and $\varphi_{\pm}$ are fields we add by hand in order to reproduce the IR SCFT that the $\CT_X$ theory reduced on a circle flows to. This situation is quite unlike what happens for the mirrors of many of the dimensional reductions of the AD theories discussed in \cite{Xie:2012hs,Xie:2013jc}} (see also the discussions in \cite{Buican:2015hsa,Buican:2017uka,Benvenuti:2017lle}).
\end{itemize}

\newsec{A remark on the Hall-Littlewood chiral ring of $\CT_X$ and the Schur sector}
In this section, we briefly discuss the Hall-Littlewood (HL) chiral ring of the $\CT_X$ theory in order to tease out some additional information about the Schur sector of the $\CT_X$ SCFT. Based on our discussion above, the HL ring is generated by the following 4D Schur operators
\begin{equation}
\mu^A_{SU(2)}\in\hat\CB_1~, \ \ \ \mu^I_{SU(3)}\in\hat\CB_1~, \ \ \ \CO^a_I\in\hat\CB_{3\over2}~,
\end{equation} 
where $A$ and $I$ are adjoint indices of $SU(2)$ and $SU(3)$ respectively, and $a$ is a fundamental index of $SU(2)$.

In \eqref{JWnon0}, we saw that $W^a_I=\chi\left[\CO^a_I\right]$ and $J^I_{SU(3)}=\chi\left[\mu^I_{SU(3)}\right]$ had a non-trivial normal-ordered product in the ${\bf2}\times{\bf1}$ channel of $SU(2)\times SU(3)$. On the other hand, as we show in Appendix \ref{HLlim}, the HL limit of the $\CT_X$ index has the following expansion
\begin{eqnarray}\label{TXHLpert}
\CI_{HL}^{\CT_X}(t,w,z_1,z_2)&=&1+(\chi_{2}+\chi_{1,1})t+\chi_{1}\chi_{1,1}t^{3\over2}+(1+\chi_{4}+\chi_{1,1}+\chi_{2}\chi_{1,1}+\chi_{2,2})t^2+\cr&+&(\chi_{1} \chi_{1,1} + \chi_{3}\chi_{1,1}+\chi_{1}\chi_{3,0} + 
\chi_{1} \chi_{0,3} + \chi_{1}\chi_{2, 2})t^{5\over2}+\CO(t^3)~.\ \ \ \ \ \ 
\end{eqnarray}
Note that, compared with the Schur index in \eqref{schurTX}, the HL index is missing a contribution of the form $\chi_1$ at $\CO(t^{5\over2})\sim\CO(q^{5\over2})$ (recall that the power of the fugacity in the HL limit of the index is also given by $h=E-R$). The only apparent explanation, given our generators and the above discussion, is that there is a relation in the HL ring of the form
\begin{equation}\label{HL52rel}
\mu_{SU(3)}^I\CO^{a}_{{3\over2}\ I}=0~.
\end{equation}
In order to reconcile this relation with \eqref{JWnon0}, we conjecture that the theory has a $\hat\CC_{{1\over2}(0,0)}$ multiplet with Schur operator, $\hat\CO^{111}_{+\dot+}$, and that this operator appears in the $SU(2)_R$ twisted OPE of the $\mu^I_{SU(3)}$ and $\CO^a_I$ operators (in the sense described in Footnote \ref{ttfootnote}) so that
\begin{equation}
\mu^I_{SU(3)}(z,\bar z)\CO^a_I(0)\supset\hat\CO_{+\dot+}^{111}(0)~.
\end{equation}
At the level of component (untwisted OPEs), we have
\begin{equation}\label{chirOPE12}
J^{4d,I}_{SU(3)}(x)\CO_I^a(0)\supset {x_{-\dot-}\over x^2}\hat\CO_{+\dot+}^{111}(0)~,
\end{equation}
where the operator on the far left of this inclusion is the $R=0$ partner of the holomorphic moment map, $\mu^I_{SU(3)}$. It is straightforward to check that such mixing is compatible with $\CN=2$ superconformal Ward identities and that therefore $\hat\CO_{+\dot+}^{111}$ maps to a normal ordered product of generators of $\chi(\CT_X)$.\footnote{Often one must use highly non-trivial superspace techniques to determine which short multiplets are allowed by $\CN=2$ superconformal symmetry to appear in the OPE of two short multiplets (e.g., see \cite{Liendo:2015ofa,Ramirez:2016lyk}). However, in our case, a more pedestrian approach suffices to show that \eqref{chirOPE12} is allowed. Indeed, we can show that such terms exist in free SCFTs. To that end, consider a free hypermultiplet
\begin{equation}
q^i=\begin{pmatrix}
    Q   \\
    \tilde Q^{\dagger} 
  \end{pmatrix}~, \ \ \ q^{\dagger i}=\tilde q^i=\begin{pmatrix}
    \tilde Q   \\
    -Q^{\dagger} 
  \end{pmatrix}~,
\end{equation}
where $i$ is an $SU(2)_R$ spin-half index. Let us construct $\hat\CB_1$ and $\hat\CB_{3\over2}$ multiplets of the form $q^{(i}\tilde q^{j)}$ and $q^{(i}q^j\tilde q^{k)}$ respectively (where \lq\lq$(\cdots)$" denotes symmetrization of the enclosed indices). This theory has a $\hat\CC_{{1\over2}(0,0)}$ multiplet with a primary of the form $\epsilon_{ij}q^i\tilde q^jq^k$. The associated Schur operator is (up to an overall normalization)
\begin{equation}\label{C12schur}
\CO_{+\dot+}^{111}\sim(\tilde Q\partial_{+\dot+}Q-Q\partial_{+\dot+}\tilde Q)Q~.
\end{equation}
We then see that \eqref{chirOPE12} is allowed by supersymmetry since a trivial computation in free field theory reveals that (at separated points)
\begin{equation}\label{3pt}
\langle(QQ^{\dagger}-\tilde Q\tilde Q^{\dagger})(x)QQ\tilde Q(y)(\tilde Q^{\dagger}\partial_{+\dot+}Q^{\dagger}-Q^{\dagger}\partial_{+\dot+}\tilde Q^{\dagger})Q^{\dagger})(0)\rangle\ne0~.
\end{equation}
}
This discussion is analogous to what happens in the OPE of moment maps in the rank one theories discussed in \cite{Beem:2013sza} (there the 2D interpretation of the corresponding OPE is that the stress tensor is a Sugawara stress tensor; in the case of the $\CT_X$ theory, the conclusion is quite different).

In the next section we will switch gears and focus on the implication of the non-vanishing Witten anomaly of $SU(2)\supset G_{\CT_X}$ for the 2D/4D correspondence of \cite{Beem:2013sza}.

\newsec{Witten's anomaly and the chiral algebra}\label{wittenanom}
One of the deepest questions in the 4D/2D correspondence of \cite{Beem:2013sza} is to understand which chiral algebras in 2D are part of a \lq\lq swampland" of theories that cannot be related to consistent (and unitary) 4D $\CN=2$ SCFTs. One example of a constraint all chiral algebras that are not part of this swampland must obey (unless they are part of the special set of chiral algebras related to a finite subset of free SCFTs in 4D with sufficiently few fields) follows from the analysis in \cite{Liendo:2015ofa}
\begin{equation}
c_{2d}\le-{22\over5}~.
\end{equation}

\begin{figure}
\begin{center}
\vskip .5cm
\begin{tikzpicture}[place/.style={circle,draw=blue!50,fill=blue!20,thick,inner sep=0pt,minimum size=6mm},transition/.style={rectangle,draw=black!50,fill=black!20,thick,inner sep=0pt,minimum size=5mm},transition2/.style={rectangle,draw=black!50,fill=red!20,thick,inner sep=0pt,minimum size=8mm},transition3/.style={rectangle,draw=black!50,fill=red!20,thick,inner sep=0pt,minimum size=10mm},auto]
\node[transition3] (1) at (1.8,0) {$\;\CT_{X}\;$};
\node[place] (2) at (3,0) [shape=circle] {$2$} edge [-] node[auto]{} (1);
\node[transition2] (3) at (4.1,0)  {$2$} edge [-] node[auto]{} (2);
\end{tikzpicture}
\caption{The above SCFT is inconsistent because of the $SU(2)$ anomaly of the $\CT_X$ theory. It would be interesting to study how this inconsistency is manifested in the chiral algebra setting.}
\label{anom}
\end{center}
\end{figure}
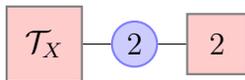

We would like to point out that another constraint chiral algebras outside the swampland must obey is that they are not related to 4D $\CN=2$ SCFTs that have a gauge symmetry with a Witten anomaly \cite{Witten:1982fp}.\footnote{However, it is conceivable that two different 4D $\CN=2$ SCFTs might have the same chiral algebra (although we are not aware of any such examples). Therefore, we cannot immediately rule out the (perhaps remote) possibility that one might have a 2D chiral algebra that is related both to a well-defined 4D SCFT and a pathological one of the type described here.} Indeed, the corresponding 4D theory is inconsistent. Interestingly, our $\CT_X$ theory allows us to construct an infinite number of pathological SCFTs by gauging the $SU(2)$ global symmetry (of course, we can also construct infinitely many conformal manifolds that are consistent and have no Witten anomaly; note that the $\CT_X$ theory on its own is also perfectly consistent since the $SU(2)$ symmetry is global).

A simple example of such a pathological theory is given in Fig. \ref{anom}. To construct this SCFT, we gauge a diagonal $SU(2)$ flavor symmetry of the $T_2$ and $\CT_X$ theories (where the $\CT_X$ contribution is the anomalous $SU(2)$ factor and not a subgroup of $SU(3)$). Using the expression for the $T_2$ index given in \cite{Lemos:2014lua} and our expression in \eqref{2d partition}, it is straightforward to verify that the naive index of the pathological theory is\footnote{We are making this statement at the naive level of operator counting. Note that the $Z_{S^1\times S^3}$ partition function (which differs from the index by certain pre-factors) may have additional pathologies.}
\begin{eqnarray}\label{wittenanomI}
\CI(q,y_1,y_2,z_1,z_2)&=&\sum_{\lambda}q^{2\lambda}P.E.\left[{2q^2\over1-q}+2q-2q^{1+\lambda}\right]{\rm ch}_{R_{\lambda}}^{SU(2)}(q,y_1){\rm ch}_{R_{\lambda}}^{SU(2)}(q,y_2)\times\cr&\times&{\rm ch}_{R_{\lambda,\lambda}}^{SU(3)}(q,z_1,z_2)~,
\end{eqnarray}
where $y_{1,2}$ are $SO(4)$ fugacties, and $z_{1,2}$ are the $SU(3)$ fugacities introduced above.

It would be interesting to understand how (or even if!) this pathology is manifested in the 2D setting. One possibility is that such chiral algebras (like the one whose vacuum character is given in \eqref{wittenanomI}) are somehow pathological (or perhaps the non-trivial representations of these chiral algebras are pathological). Another possibility is that the chiral algebras and their modules are perfectly consistent at the level of 2D QFT but still detect the pathology of the 4D theory. While we have not fully investigated this question, we suspect the latter possibility holds (we should also note that, in principle, it could be that the chiral algebra and its representations are perfectly consistent and also do not detect the 4D pathology). We hope to return to this question soon.\footnote{It may be possible to use some of the theories described in \cite{Argyres:2007tq,Argyres:2016xmc} to study this question as well.}

\newsec{Conclusions and open questions}
Using very little data, we found the Schur index and chiral algebra of the exotic isolated irreducible SCFT, $\CT_X$,\footnote{Note that this chiral algebra lies outside the classes of AD chiral algebras considered in the literature before (e.g., see \cite{Buican:2015ina,Cordova:2015nma,Cecotti:2015lab,Buican:2016arp,Creutzig:2017qyf,Song:2017oew}).} that emerges in the simplest AD generalization of Argyres-Seiberg duality. Moreover, we saw this theory has a remarkable resemblance to its cousin $T_N$ theories (although its chiral algebra is even simpler) and that, like the other component theories of the duality described in \cite{Buican:2014hfa}, the $\CT_X$ Schur index is intimately related to the index of free fields (even though the theory itself is strongly interacting). As a result of this study, we found a more pleasing place for the duality described in \cite{Buican:2014hfa} in the landscape of $\CN=2$ dualities.

Our work raises many open questions. Among them are the following:
\begin{itemize}
\item{Is there a deeper relation between the $\CT_X$ SCFT and the $T_N$ theories? We saw the Schur indices were closely related. What about more general limits of the index? Is there a family of $\CT_X$ theories arising from $\CN=2$ $S$-dualities that are close cousins of the $T_N$ theories?}
\item{Is there an explanation for why all the component theories in the duality we considered have Schur indices that are so closely related to those of free fields (perhaps generalizing the reasoning in \cite{kac2017remark})? Could this be some interesting manifestation of modularity in disguise?}
\item{We saw that our indices are naturally written in terms of AKM characters. Is there a form of the index that is more natural from a TFT perspective (perhaps generalizing \cite{Buican:2015ina,Song:2015wta,Buican:2017uka,Song:2017oew})?}
\item{We know that the $\CT_{3,{3\over2}}$ theory has a class $\CS$ description (using the results in \cite{Xie:2012hs}). Does the $\CT_X$ theory have such a description? Could the TFT description of the index shed some light on this question?}
\item{If the $\CT_X$ theory has a class $\CS$ description, is there a geometrical way to encode the presence of the Witten anomaly in a puncture?}
\item{This theory lacks $\CD\oplus\bar\CD$ operators in its HL ring. Is this absence a clue for the appropriate way to think about the topology of the Riemann surface in this case (again, assuming the theory is class $\CS$)? See \cite{Xie:2017vaf} for some recent ideas on the topology that is naturally associated with AD theories.}
\item{The $\CT_X$ chiral algebra has only bosonic operators. Is this part of some larger pattern for isolated $1<\CN<3$ SCFTs?}
\item{Our theory has $SU(3)\times SU(2)\times U(1)$ flavor symmetry (when viewed as an $\CN=1$ theory). We are not aware of another way to find this symmetry group in string or field theory from a minimality condition (recall that in our case, this symmetry emerges from requiring that we study the minimal generalization of Argyres-Seiberg duality to $\CN=2$ SCFTs with non-integer chiral primaries). Can the minimality we are discussing be made more precise so that one can find this SCFT using the conformal bootstrap (perhaps, in light of \eqref{HL52rel} and \eqref{JWnon0}, it will be useful to study the $\langle J^I W^a_K J^L W^b_M\rangle$ four-point function)? What if we gauge the flavor symmetry--can this SCFT act as a hidden sector for beyond the standard model physics (since the $U(1)$ is not asymptotically free, this gauged theory can, at best, be part of an effective field theory)?}
\item{Is it possible to make contact with a generalization of \cite{Gadde:2015xta,Maruyoshi:2016tqk} to the case at hand?}
\item{Can we find a manifestation of the 4D Witten anomaly for the (inconsistent) SCFT in Fig. \ref{anom} in the corresponding 2D chiral algebra (as discussed in Sec. \ref{wittenanom})?}
\item{As a final amusing note, it is interesting to observe that the expression in \eqref{2d partition} makes it rather trivial to write down simple formulae for the indices of conformal manifolds built out of $\CT_X$ theories (as in the case of the $T_N$ theories). For typical conformal manifolds built out of AD theories (e.g., as in the case of the $(A_N, A_M)$ conformal manifolds studied in \cite{Buican:2017uka}), this procedure is considerably more complicated.}
\end{itemize}

\ack{We would like to thank S.~Benvenuti, S.~Giacomelli, C.~Papageorgakis, and D.~Xie for discussions on this and related topics. M.~B. would like to thank the ICTP and SISSA theory groups for stimulating discussions and hospitality during the completion of this work. M. B. would also like to thank the Aspen Center for Physics (NSF grant \#PHY-1066293) and Perimeter Institute (supported by the Government of Canada and the Province of Ontario) for hosting interesting workshops during the course of this research. M.~B.'s research is partially supported by the Royal Society under the grant \lq\lq New Constraints and Phenomena in Quantum Field Theory." Z.~L. is supported by a Queen Mary University of London PhD studentship. T.~N.'s research was partially supported by JSPS Grant-in-Aid for Scientific Research (B) No. 16H03979. }

\newpage

\begin{appendices}
\section{Proof of the XYY formula}\label{app:PEproof}
In this appendix we review the fact that the conjectured XYY formula for the Schur index of the $(A_1,D_4)$ theory \cite{Xie:2016evu} reproduced in \eqref{XYYform} can be proven using Theorem 5.5 of \cite{Creutzig:2017qyf} (in fact, this result follows directly from (11) of \cite{kac2017remark}).\footnote{Note that the authors of \cite{kac2017remark} also demonstrate more general conjectures \cite{Xie:2016evu} for theories closely related to the $(A_1, D_4)$ SCFT.} 

To that end, we start with the XYY formula
\begin{eqnarray}\label{XYYform2}
\CI_{(A_1, D_4)}(q,a,b)&=&\mathrm{P.E.}\left[\frac{q}{1-q^{2}}\,\chi_{\text{Adj}}^{SU(3)}(a,b)\right]\cr&=&\mathrm{P.E.}\left[\frac{q}{1-q^{2}}(2+\frac{1}{a^{2}b}+\frac{1}{ab^{2}}+\frac{a}{b}+\frac{b}{a}+a^{2}b+ab^{2})\right]~.
\end{eqnarray}
Expanding the plethystic exponentials, we obtain
\begin{equation} \label{PEconv}
\mathrm{P.E.}\left[\frac{a}{1-b}\right]=\prod_{i=0}^{\infty}{1\over1-ab^i}~,
\end{equation}
and we can then rewrite \eqref{XYYform2} as
\begin{align}\label{rewriteXYY}
\mathrm{P.E.}\left[\frac{q}{1-q^{2}}\,\chi_{\text{Adj}}^{SU(3)}(a,b)\right]=&\prod_{n=0}^{\infty}\frac{1}{(1-q^{2n+1})^{2}}\frac{1}{(1-\frac{1}{a^2b}q^{2n+1})}\frac{1}{(1-\frac{1}{ab^2}q^{2n+1})}\frac{1}{(1-\frac{a}{b}q^{2n+1})}\times\nonumber\\ \times&\frac{1}{(1-\frac{b}{a}q^{2n+1})}\frac{1}{(1-a^{2}bq^{2n+1})}\frac{1}{(1-ab^{2}q^{2n+1})}~.
\end{align}
It is then straightforward to show that \eqref{rewriteXYY} becomes the Schur index of the $(A_1,D_4)$ SCFT given by Theorem 5.5 of \cite{Creutzig:2017qyf} (setting $p=2$ and with the $q^{1/3}$ prefactor stripped off)
\begin{align}\label{CA1D4}
\CI_{(A_1,D_4)}(q,x,y)=&\prod_{n=0}^{\infty}\frac{\left(1-y^{2}q^{2(n+1)}\right)\left(1-q^{2(n+1)}\right)^{2}\left(1-y^{-2}q^{2(n+1)}\right)}{\left(1-y^{2}q^{n+1}\right)\left(1-q^{n+1}\right)^{2}\left(1-y^{-2}q^{n+1}\right)\left(1-x y q^{2\left(n+\frac{1}{2}\right)}\right)}\times \nonumber \\
&\times\frac{1}{\left(1-x^{-1} y q^{2\left(n+\frac{1}{2}\right)}\right)\left(1-x y^{-1} q^{2\left(n+\frac{1}{2}\right)}\right)\left(1-x^{-1} y^{-1} q^{2\left(n+\frac{1}{2}\right)}\right)}\nonumber\\&=\prod_{n=0}^{\infty}\frac{1}{\left(1-q^{2n+1}\right)^{2}\left(1-y^{\pm 2}q^{2n+1}\right)\left(1-x^{\pm} y^{\pm} q^{2n+1}\right)}~,
\end{align}
under the fugacity map
\begin{equation}\label{fugacitymap}
a=y\,x^{1/3} \qquad b=y^{-1}\,x^{1/3}~.
\end{equation}
The relation in \eqref{fugacitymap} corresponds to the decomposition of the $SU(3)$ fugacities into fugacities of $SU(2)\times U(1)$. Before concluding, note that, as in \eqref{repeat}, the \lq\lq$\pm$" superscripts in \eqref{CA1D4} are understood as a product over each sign, e.g.
\begin{equation}
{1\over1-y^{\pm2}q^{2n+1}}\equiv {1\over1-y^{2}q^{2n+1}}{1\over1-y^{-2}q^{2n+1}}~.
\end{equation}

\section{Details of the Inversion Formula}\label{inversionap}
In this appendix we find an integral expression for the superconformal index of the $\CT_{3,{3\over2}}$ theory in the Schur limit by employing the inversion theorem proved in \cite{spiridonov2006inversions}. Our use of the inversion theorem is similar to its use in the case of the $E_6$ SCFT by the authors of \cite{Gadde:2010te}, but there are some technical differences here since our $SU(2)$ duality frame in Fig. \ref{quiver2} has, in addition to the $\CT_{3,{3\over2}}$ theory, a strongly interacting $(A_1, D_4)$ SCFT instead of a pair of hypermultiplets as in the $E_6$ case. Nonetheless, we will argue that, using the results reviewed in Appendix \ref{app:PEproof} and an argument about analytic properties of the index, we can invert the gauge integral of the index in the $SU(2)$ duality frame.

In order to find the index in the two duality frames we need the index of the basic building blocks in Figs. \ref{quiver1} and \ref{quiver2}. To that end, the single letter index of the $\mathcal{N}=2$ vector multiplet (transforming in the adjoint of the gauge group) and half-hypermultiplet (transforming in representation $R$ of the combined gauge and flavor groups) can be found in \cite{Gadde:2011uv} (whose labelling conventions for fugacities we follow). Here we reproduce these indices in the Schur limit
\begin{eqnarray}
\CI_{\rm vect}(q,{\bf x})&=&-\frac{2q}{1-q}\chi_{\text{adj}}({\bf x})~,\cr
\CI_{\frac{1}{2}H}(q,{\bf x},{\bf z})&=&\frac{\sqrt{q}}{1-q}\,\chi_{R}({\bf x},{\bf z})~.
\end{eqnarray}
We can \lq\lq glue" these indices along with the index of the $(A_1,D_4)$ SCFT given in \eqref{XYYform} by integrating their product over the Haar measure of the diagonal subgroup we are gauging.

We start with the $SU(3)$ side of the duality where we are gauging the diagonal part of the $SU(3)$ flavor symmetries of the two $(A_1,D_4)$ theories along with 3 fundamental hypermultiplets as in Fig. \ref{quiver1}. The latter degrees of freedom supply the $U(3)$ symmetry, which is decomposed as $U(3)=SU(3)_z\otimes U(1)_s$. The index on this side of the duality is then given by
\begin{align}\label{su3sideapp}
&\CI_{SU(3)}(q,s,z_1,z_2)=\frac{(q;q)^4}{6}\oint_{\mathbb{T}^2} \prod_{k=1}^{2}\frac{dx_k}{2 \pi i x_k} \prod_{i\neq j}(x_i-x_j) \left(q \frac{x_i}{x_j};q\right)^2 \times \nonumber\\
& \times \mathrm{P.E.}\left[\frac{q}{1-q^{2}}\,\chi_{\text{Adj}}^{SU(3)}(x_1,x_2)\right]^2 \prod_{i,j} \left( \sqrt{q} \left(\frac{z_{j}s^{1/3}}{x_{i}} \right)^{\pm};q \right)^{-1}~,
\end{align}
where $\mathbb{T}$ is the positively oriented unit circle, $\prod_{k=1}^{2}\frac{dx_k}{2 \pi i x_k} \frac{1}{3!}\prod_{i\neq j}(x_i-x_j)$ is the Haar measure of $SU(3)$, and the $x_i$ ($i = 1,2,3$) satisfy the constraint $\prod_{i=1}^3x_i=1$. We can rewrite \eqref{su3sideapp} slightly using elementary computations described in appendix \ref{app:PEproof}
\begin{equation} \label{rew}
\mathrm{P.E.}\left[\frac{q}{1-q^{2}}\,\chi_{\text{adj}}^{SU(3)}(x_1,x_2)\right]= (q;q^2)^{-2}\prod_{i\neq j}\left(q\frac{x_i}{x_j};q^2\right)^{-1}~, \ \ \ x_3=x_1^{-1}x_2^{-1}~.
\end{equation}
Substituting \eqref{rew} into \eqref{su3sideapp} and performing some simplifications yields the following explicit formula
\begin{align} \label{SU(3)side}
\CI_{SU(3)}(q,s,z_1,z_2)=\frac{(q^2;q^2)^4}{6}\oint_{\mathbb{T}^2} \prod_{i=1}^{2}\frac{dx_i}{2 \pi i x_i} \prod_{i\neq j}(x_i-x_j) \left(q^2 \frac{x_i}{x_j};q^2\right)^2 \prod_{i,j} \left( \sqrt{q} \left(\frac{z_{j}s^{1/3}}{x_{i}} \right)^{\pm};q \right)^{-1}~.
\end{align}
Since the index is invariant under duality transformations, \eqref{SU(3)side} has to equal the index on the $SU(2)$ side of the duality where we are gauging the diagonal $SU(2)_e$ of the $(A_1,D_4)$ and $\CT_{3,{3\over2}}$ theories as in Fig. \ref{quiver2}. We can write the index in this duality frame as
\begin{align}
\CI_{SU(2)}(q,s,z_1,z_2)=\frac{(q;q)^2}{2} \oint_\mathbb{T} \frac{de}{2 \pi i e}(e^{\pm 2}q;q)^2 &(1-e^{\pm 2})\times\nonumber\\ \times\mathrm{P.E.}\left[\frac{q}{1-q^2}\,\chi_{\text{adj}}^{SU(3)}\left(es^{\frac{1}{3}},e^{-1}s^{\frac{1}{3}},s^{-\frac{2}{3}}\right)\right] &\CI_{\mathrm{\CT_{3,{3\over2}}}}(q,e,z_1,z_2)~,
\end{align}
where $\frac{de}{2 \pi i e}\frac{1}{2}(e-e^{-1})(e^{-1}-e)$ is the Haar measure of $SU(2)$.
Rewriting the plethystic exponential as in \eqref{rew} and performing some simplifications leads to
\begin{align}
\CI_{SU(2)}(q,s,z_1,z_2)=\frac{(q^2;q^2)^2}{2} \oint_\mathbb{T} \frac{de}{2 \pi i e}\frac{(e^{\pm 2}q;q)(e^{\pm 2};q^2) \CI_{\mathrm{\CT_{3,{3\over2}}}}(q,e,z_1,z_2)}{(qs^{\pm} e^{\pm};q^2)} ~.
\end{align}
Finally, to make contact with the inversion theorem, we replace $q\rightarrow \sqrt{q}$ 
\begin{eqnarray} \label{comp}
\CI_{SU(2)}(q,s,z_1,z_2)|_{q\rightarrow \sqrt{q}}=\frac{(q;q)^2}{2} \oint_\mathbb{T} \frac{de}{2 \pi i e}\frac{(e^{\pm 2};q) }{(\sqrt{q}s^{\pm} e^{\pm};q)} (e^{\pm 2}\sqrt{q};\sqrt{q})\CI_{\mathrm{\CT_{3,{3\over2}}}}(q,e,z_1,z_2)|_{q\rightarrow \sqrt{q}}~.\ \
\end{eqnarray}
Now we will explain how to use the inversion theorem in order to extract $\CI_{\CT_{3,{3\over2}}}$ from this equation. Extracting $\CI_{\CT_{3,{3\over2}}}$ is highly non-trivial since it is not at all obvious why \eqref{comp} preserves all the information about this quantity.

\subsection{Inversion Theorem}
This subsection closely follows Appendix B of \cite{Gadde:2010te}. The input to the inversion theorem of \cite{spiridonov2006inversions} is the following type of contour integral
\begin{align} \label{invin}
\hat{f}(w)=\kappa \oint_{C_{w}}\frac{ds}{2 \pi i s}\delta(s,w;T^{-1},p,q)f(s)~,
\end{align}
where $\kappa=\frac{1}{2}(p;p)(q;q)$, $w$ is on the unit circle, and the integral kernel is defined as
\begin{equation} \label{delta}
\delta(s,w;T,p,q) \equiv \frac{\Gamma(T s^{\pm 1}w^{\pm 1};p,q)}{\Gamma(T^{2};p,q)\Gamma (s^{\pm 2};p,q)}~.
\end{equation}
In \eqref{delta}, $T$ is a function of $p,q,t\in\mathbb{C}$ satisfying
\begin{equation}\label{eq:cond}
\mathrm{max}\,\left(|p|,|q|\right)<|T|^2<1~,
\end{equation}
$\Gamma(z;p,q)$ is defined as
\begin{equation}\label{Gammadef}
\Gamma(z;p,q)\equiv\prod_{j,k\ge0}{1-z^{-1}p^{j+1}q^{k+1}\over1-zp^jq^k}~,
\end{equation}
and $f(s)\equiv f(s,p,q,t)$ is a function that is holomorphic in the annulus 
\begin{equation}
\mathbb{A}=\{|T|-\varepsilon<|s|<|T|^{-1}+\varepsilon\}~,
\end{equation}
for small but finite $\varepsilon>0$ and also satisfies 
\begin{equation}\label{scond}
f(s)=f(s^{-1})~.
\end{equation}
The contour $C_{w}=C_{w}^{-1}$ lies in the annulus $\mathbb{A}$
with the points $T^{-1}w^{\pm}$ in its interior (and therefore the points $Tw^{\pm}$ in its exterior). If these conditions are all satisfied, then the inversion theorem states that $f$ can be recovered from the contour integral
\begin{align} \label{invout}
f(s)=\kappa \oint_{\mathbb{T}}\frac{de}{2 \pi i e} \delta(e,s;T,p,q)\hat{f}(e)~.
\end{align}

As first applied to the index in \cite{Gadde:2010te}, this inversion theorem is used as follows. First, one finds a representation of the conformal manifold index that is of the form of the RHS of \eqref{invout}. In particular, $\hat f(e)$ should contain the index of the isolated SCFT (the $E_6$ theory in \cite{Gadde:2010te} or the $\CT_{3,{3\over2}}$ SCFT in the case at hand) we wish to determine. One then makes an analytic assumption that $\hat f(e)$ can be written as in \eqref{invin} for some function $f(s)$ satisfying \eqref{scond} while being analytic in the annulus, $\mathbb{A}$. Then, the inversion theorem implies that $f(s)$ is the index of the conformal manifold. However, in general, one is not guaranteed that the analytic assumption described above holds.\footnote{Therefore, the authors of  \cite{Gadde:2010te} performed many non-trivial consistency checks of this procedure in the $E_6$ case. Our results in the main text can be viewed as highly non-trivial consistency checks of this procedure for the $\CT_{3,{3\over2}}$ SCFT.}

As a result, to apply this theorem in our case, we first need to choose $\hat f(e)$ in \eqref{invout} so that \eqref{invout} coincides with \eqref{comp}. To that end, using 
\begin{equation}
\Gamma(z;p,q)=\mathrm{P.E.}\left[ \frac{z-pq/z}{(1-p)(1-q)}\right]~,
\end{equation}
and \eqref{PEconv} one finds that the \lq\lq delta function" in \eqref{invout} satisfies (for our choice of $T$ discussed below)
\begin{align}\label{deltadefn}
\delta(e,s;T,p,q)= \frac{(T^2;q)(e^{\pm 2};q)}{(T e^{\pm} s^{\pm};q)}\tilde\delta(e;T,p,q)~,
\end{align}
where $\tilde\delta(e;T,p,q)$ contains $p$-dependent terms. By comparing \eqref{comp} with \eqref{invout}, one can see that if we choose $T=\sqrt{q}$ and
\begin{equation} \label{sep}
\hat{f}(e)=(e^{\pm 2}\sqrt{q};\sqrt{q})\times (e^{\pm2}p;p)^{-1}\times\CI_{\mathrm{\CT_{3,{3\over2}}}}(q,e,z_1,z_2)|_{q\rightarrow \sqrt{q}}~,
\end{equation} 
the two expressions coincide.

However, there is an additional wrinkle in our application of the inversion theorem relative to the $E_6$ case in \cite{Gadde:2010te}. Indeed, under the analytic assumption described in the paragraph below \eqref{invout}, we have
\begin{eqnarray}\label{inversion2}
(w^{\pm 2} \sqrt{q};\sqrt{q})\times(w^{\pm2}p;p)^{-1}&\times&\CI_{\mathrm{\CT_{3,{3\over2}}}}(q,w,z_1,z_2)|_{q\rightarrow\sqrt{q}}=\frac{(q;q)(p;p)}{2}\times\oint_{C_w}\frac{ds}{2 \pi i s} \frac{(\frac{1}{q};q)(s^{\pm 2};q)}{(\frac{1}{\sqrt{q}}s^{\pm}w^{\pm};q)}\times\cr&\times&\tilde\delta(s,w;{1\over\sqrt{q}},p,q)\times\CI_{SU(3)}(q,s,z_1,z_2)|_{q\rightarrow \sqrt{q}}~,
\end{eqnarray}
where, as in \eqref{deltadefn}, we have separated $\delta$ into a $p$-independent part and a $p$-dependent part, $\tilde\delta$. While the $p$-dependence in \eqref{inversion2} can be cancelled so that
\begin{eqnarray}\label{inversion3}
(w^{\pm 2} \sqrt{q};\sqrt{q})&\times&\CI_{\mathrm{\CT_{3,{3\over2}}}}(q,w,z_1,z_2)|_{q\rightarrow\sqrt{q}}=\frac{(q;q)}{2}\times\oint_{C_w}\frac{ds}{2 \pi i s} \frac{(\frac{1}{q};q)(s^{\pm 2};q)}{(\frac{1}{\sqrt{q}}s^{\pm}w^{\pm};q)}\times\cr&\times&\CI_{SU(3)}(q,s,z_1,z_2)|_{q\rightarrow \sqrt{q}}~,
\end{eqnarray}
the condition \eqref{eq:cond} fails for $T=\sqrt{q}$, and $\delta(s,w;{1\over\sqrt{q}};p,q)=0$ (since $({1\over q};q)=0$). Therefore, the RHS of \eqref{inversion3} vanishes.\footnote{A similar situation occurs in the $E_6$ example of \cite{Gadde:2010te} if one first takes the Schur limit and then performs the integration.}

To get a more sensible answer, we can consider taking $T=\sqrt{q}(1+\varepsilon')$ for $\varepsilon'\ll1$. In this case, we have
\begin{equation}\label{zerofact}
\left({1\over q};q\right)\to\left({1-2\varepsilon'\over q};q\right)\ne0~,
\end{equation}
and the expression on the RHS of \eqref{inversion3} is non-vanishing since it becomes 
\begin{equation}\label{inversion4}
\frac{(q;q)}{2}\times\oint_{C_w}\frac{ds}{2 \pi i s} \frac{(\frac{1-2\varepsilon'}{q};q)(s^{\pm 2};q)}{(\frac{1-\varepsilon'}{\sqrt{q}}s^{\pm}w^{\pm};q)}\times\CI_{SU(3)}(q,s,z_1,z_2)|_{q\rightarrow \sqrt{q}}~.
\end{equation}
In particular, note that the double poles at $s=T^{-1}w^{\pm1}$ and $s=qT^{-1}w^{\pm1}$ in \eqref{inversion3} are resolved into eight single poles in \eqref{inversion4} with one of each pair still taken to be in the integration contour (for a total of four) and a factor of $\varepsilon'^{-1}$ from the residues that cancels the factor of $\varepsilon'$ arising from \eqref{zerofact} (all other contributions will be parametrically smaller in $\varepsilon'$). Taking the $\varepsilon'\to0$ limit then gives us a prescription for computing the Schur index with
\begin{eqnarray}\label{prescription}
(w^{\pm 2} \sqrt{q};\sqrt{q})\times\CI_{\mathrm{\CT_{3,{3\over2}}}}(q,w,z_1,z_2)|_{q\rightarrow\sqrt{q}}&=&\lim_{\varepsilon' \rightarrow 0}\frac{(q;q)}{2}\times\oint_{C_w}\frac{ds}{2 \pi i s} \frac{(\frac{1-2\varepsilon'}{q};q)(s^{\pm 2};q)}{(\frac{1-\varepsilon'}{\sqrt{q}}s^{\pm}w^{\pm};q)}\times\cr&\times&\CI_{SU(3)}(q,s,z_1,z_2)|_{q\rightarrow \sqrt{q}}~.
\end{eqnarray}
The contour integration around an infinite number of poles thus reduces to the residues of just four poles whose contribution gives us the simple expression
\begin{equation}
\CI_{\CT_{3,{3\over2}}}(q,w,z_1,z_2)=\frac{1}{(w^{\pm 2} q;q)}\left[ \frac{1}{1-w^2}\CI_{SU(3)}(q,wq,z_1,z_2)+\frac{w^2}{w^2-1}\CI_{SU(3)}(q,\frac{q}{w},z_1,z_2)\right]~.
\end{equation}

We can justify the above discussion a posteriori by noting that the non-trivial checks in the main text strongly suggest that \eqref{prescription} is a consistent prescription. While a similar procedure works for the Schur index of the $E_6$ SCFT discussed in \cite{Gadde:2010te}, our case at hand is somewhat more special. Indeed, we used the fact that the $(A_1, D_4)$ SCFT has a Schur index whose $s$ dependence (after taking $q\to\sqrt{q}$) in \eqref{comp} is the same as for $\delta(e,s;\sqrt{q},p,q)$. On the other hand, when we take $T\to\sqrt{q}(1+\varepsilon')$, we do not necessarily expect that the $(A_1, D_4)$ SCFT has a limit of the index whose $s$ dependence matches the $s$ dependence in $\delta(e,s;\sqrt{q}(1+\varepsilon'),p,q)$ to all orders in $\varepsilon'$. However, the $\CO(\epsilon')$ resolution of the double poles into single poles described above should correspond to a shift in the fugacities of the index so that previously degenerate contributions from sets of operators are no longer degenerate (this statement is quite natural since generic single letter contributions to the index will be shifted at $\CO(\epsilon')$ if we identify $T$ with a fugacity) and that higher-order differences with respect to $\delta(e,s;\sqrt{q}(1+\varepsilon'),p,q)$ do not affect the validity of our computation in the limit of small $\varepsilon'$.

\section{$q\rightarrow1$ and $S^3$ partition function}\label{S3partfn}
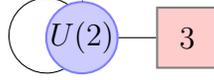
\begin{figure}
\begin{center}
\vskip .5cm
\begin{tikzpicture}[place/.style={circle,draw=blue!50,fill=blue!20,thick,inner sep=0pt,minimum size=6mm},transition2/.style={rectangle,draw=black!50,fill=red!20,thick,inner sep=0pt,minimum size=8mm},auto]
\draw [black] (3,.2) arc [radius=0.5, start angle=20, end angle= 340];
\node[place] (2) at (3,0) [shape=circle] {$U(2)$} edge [-] node[auto]{} (2);
\node[transition2] (3) at (4.4,0)  {$3$} edge [-] node[auto]{} (2);
\end{tikzpicture}
\caption{The quiver diagram describing the $S^1$ reduction of the $\CT_{3,{3\over2}}$ theory (it is mirror to the mirror in Fig. \ref{quiver3}). The closed loop beginning and ending at the $U(2)$ node denotes an adjoint hypermultiplet of $U(2)$.}
\label{quiver5}
\end{center}
\end{figure}

The superconformal index can alternatively be viewed as a partition function on $S^3\times S^1$. Moreover, the fugacity $q=e^{-\beta}$ introduced in the main text controls the relative radii of the $S^3$ and $S^1$ factors. In particular, in the $\beta\to0$ limit, the $S^1$ factor shrinks relative to the $S^3$ factor and, up to divergent terms, we expect the index to reduce to the $S^3$ partition function, $Z_{S^3}$.

In the limit of $\beta\to0$, our expression for the $\CT_X$ index in \eqref{2d partition} can be described by the rules in \cite{Nishioka:2011dq}. In particular, the sum over $\lambda$ is replaced by an integral on $m$, where
\begin{equation}
\lambda=-\frac{2\pi m}{\beta}~,
\end{equation} 
and the group fugacities are $w=e^{-i\beta \zeta}$, $z_i=e^{-i\beta \zeta_i}$. We drop group fugacity independent factors in \eqref{2d partition} and only work to leading order in $\beta$. The $\beta \rightarrow 0$ limit of the remaining quantities are given by the following dictionary \cite{Nishioka:2011dq}
\begin{eqnarray}\label{dictionary}
\mathrm{P.E.}[-2q^{\lambda+1}]&\rightarrow& (1-e^{2 \pi m})^2~,\cr
\dim_q R_{\lambda}^{SU(2)}&\rightarrow& \sinh(\pi m)~,\cr
\dim_q R_{\lambda,\lambda}^{SU(3)} &\rightarrow& \sinh(2 \pi m) \sinh^2(\pi m)~,\cr
\mathrm{P.E.}\left[\frac{q}{1-q}\chi_{adj}\right] &\rightarrow&\prod_{j<k}\frac{(\zeta_j-\zeta_k)}{\sinh \pi (\zeta_j-\zeta_k)}~,\cr
\chi_{R_\lambda}^{SU(2)}(w) &\rightarrow& \frac{\sin (2 \pi m \zeta)}{\zeta}~,\cr
\chi_{R_{\lambda,\lambda}}^{SU(3)}(z_1,z_2,z_3)&\rightarrow& \frac{\sin \pi m(\zeta_1-\zeta_2)\sin \pi m(2\zeta_1+\zeta_2)\sin \pi m(2\zeta_2+\zeta_1)}{(\zeta_1-\zeta_2)(2\zeta_1+\zeta_2)(2\zeta_2+\zeta_1)}~.
\end{eqnarray}
Using \eqref{dictionary} and replacing the sum over $\lambda$ with an integral over $m$, the $\beta\rightarrow 0$ limit of \eqref{2d partition} becomes
\begin{align}\label{q to 1}
\int_{-\infty}^{\infty} \frac{d m}{\sinh 2 \pi m \sinh \pi m} \frac{\sin \pi m(\zeta_1-\zeta_2)\sin \pi m(2\zeta_1+\zeta_2)\sin \pi m(2\zeta_2+\zeta_1)}{\sinh \pi (\zeta_1-\zeta_2)\sinh \pi (2\zeta_1+\zeta_2)\sinh \pi (2\zeta_2+\zeta_1)} \frac{\sin 2\pi m \zeta}{\sinh 2 \pi \zeta}~.
\end{align}
One can integrate \eqref{q to 1} by turning it into a contour integral and using the residue theorem. The result is the following
\begin{align}\label{intresult}
& \frac{1}{32}\mathrm{sech\,\pi \zeta}\,(2\mathrm{csch}\,\pi  (\zeta_1-\zeta_2)\, \mathrm{csch}\,\pi  (\zeta_1+2 \zeta_2)\, \mathrm{sech}\,\pi  (\zeta -2 \zeta_1-\zeta_2)\, \mathrm{sech}\,\pi  (\zeta +2 \zeta_1+\zeta_2)\nonumber\\
&\qquad-\mathrm{csch}\,\pi  (2 \zeta_1+\zeta_2) \,\mathrm{csch}\,\pi  (\zeta_1+2 \zeta_2)\, \mathrm{sech}\,\pi  (\zeta +\zeta_1-\zeta_2)\, \mathrm{sech}\,\pi  (\zeta -\zeta_1+\zeta_2) \nonumber \\
&\qquad-\text{csch}\,\pi  \zeta \,\text{csch}\, \pi  \left(\zeta _1-\zeta _2\right)\, \text{csch}\,\pi  \left(\zeta _1+2 \zeta _2\right)\nonumber \\
&\qquad \qquad\times (\left(2 \zeta +3 \zeta _1+5 \zeta _2\right) \text{sech}\,\pi  \left(\zeta -\zeta _1-2 \zeta _2\right)\, \text{sech}\,\pi  \left(\zeta +\zeta _1-\zeta _2\right)\nonumber\\
&\qquad\qquad-\left(4 \zeta +3 \zeta _1+5 \zeta _2\right) \,\text{sech}\,\pi  \left(\zeta -\zeta _1+\zeta _2\right)\, \text{sech}\,\pi  \left(\zeta +\zeta _1+2 \zeta _2\right))\nonumber\\
&\qquad-\frac{1}{2}\text{csch}\,\pi  \zeta \,\text{csch}\,\pi  \left(2 \zeta _1+\zeta _2\right)\,\text{csch}\,\pi  \left(\zeta _1+2 \zeta _2\right)\nonumber \\
&\qquad\qquad\times (\left(4 \zeta +\zeta _1+\zeta _2\right) \text{sech}\,\pi  \left(\zeta -\zeta _1-2 \zeta _2\right) \text{sech}\,\pi  \left(\zeta -2 \zeta _1-\zeta _2\right)\nonumber\\
&\qquad\qquad-\left(2 \zeta +\zeta _1+\zeta _2\right) \text{sech}\, \pi  \left(\zeta +2 \zeta _1+\zeta _2\right) \text{sech}\,\pi  \left(\zeta +\zeta _1+2 \zeta _2\right))\nonumber \\
&+(\zeta_1 \leftrightarrow \zeta_2)~.
\end{align}
This answer can then be compared with the partition function of the $S^1$ reduction of $\CT_{X}$ or of the mirror theory in Fig. \ref{quiver3}. The direct $S^1$ reduction of $\CT_{3,{3\over2}}$ is described by an $\CN=4$ $U(2)$ gauge theory whose Lagrangian quiver is illustrated in Fig. \ref{quiver5} \cite{Cremonesi:2014xha}. Once we decouple the contribution of the $SU(2)$ gauge singlet part of the adjoint hypermultiplet, $\frac{1}{\cosh \pi m’}$, which is the 3D descendant of the decoupled hyper of $\CT_{3,{3\over2}}$ we can write down the partition function of the 3D reduction of $\CT_{X}$ \cite{Hama:2010av} \cite{Benvenuti:2011ga} 
\begin{eqnarray}\label{S1 reduction}
Z_{S^3}^{\rm quiver}&=&{1\over2}\int_{-\infty}^{\infty} dx_1 dx_2 \frac{\sinh^2(\pi(x_1 - x_2)) e^{2\pi i \eta(x_1 + x_2)}}{\cosh\pi(x_1 - x_2 - m’) \cosh\pi(x_2 - x_1 - m’)} \cr&\times&\frac{1}{\cosh\pi m'\cosh\pi(x_1 - m_1)  \cosh\pi(x_2 - m_1) \cosh\pi(x_1 - m_2)}\cr&\times&\frac{1}{\cosh\pi(x_2 - m_2) \cosh\pi(x_1 + m_1 + m_2) \cosh\pi(x_2 + m_1 + m_2)}~.
\end{eqnarray}
This integral can be evaluated similary to \eqref{q to 1} with the same result (up to an unimportant overall constant and after using the map $\zeta\to m'$, $\zeta_i\to m_i$) as in \eqref{intresult} (again, a similar statement holds for the partition function of the mirror in Fig. \ref{quiver3}, which involves six integrations and for which one should use the fugacity map in \eqref{fmap2}).

\section{The Hall-Littlewood index of $\CT_X$}\label{HLlim}
In this appendix, we derive the HL index in \eqref{TXHLpert}. In the language of \cite{Dolan:2002zh}, the HL operators are a subset of the Shur operators described around \eqref{multtypes} and are of type $\hat\CB_R$ and $\CD_{R(0,j_2)}\oplus\bar\CD_{R(j_1,0)}$ (see Sec. \ref{2D4Dcorr} for more details). In this section we merely note that they contribute to a limit of the superconformal index described in \cite{Gadde:2011uv} where their contributions are of the form $t^{E-R}$ where $t$ is the HL superconformal fugacity (this limit of the index also detects flavor symmetries).

When a 4D $\CN=2$ theory is put on a circle, we can often compute the HL limit of the index from the 3D $\CN=4$ Higgs branch Hilbert series provided the compactification is sufficiently well-behaved. Equivalently, mirror symmetry allows us to compute the HL limit of the 4D theory from the Coulomb branch Hilbert series of the mirror theory.

Indeed, we can try to compute $\CI_{HL}^{\CT_X}$ by first computing $\CI_{HL}^{\CT_{3,{3\over2}}}$ from the 3D mirror gauge theory that follows from the rules reproduced in Fig. \ref{quiver3} and described in \cite{Xie:2012hs}.\footnote{Note that we found substantial evidence in favor of this proposed quiver in the main body of the text.} Using the results in \cite{Cremonesi:2013lqa}, we can write this index as follows
\begin{equation}\label{3DT332}
\CI_{HL}^{\CT_{3,{3\over2}}}(t)={1\over(1-t)^3}\sum_{a_1, a_{A,i},a_{B,i}\in\Gamma^*_{\hat G}/\mathcal{W}_{\hat G}}\zeta_A^{a_{A,1}+a_{A,2}}\zeta_B^{a_{B,1}+a_{B,2}}\zeta_C^{a_{C,1}+a_{C,2}}\cdot P(a_{A,i},a_{B,i},a_{C,i})\cdot t^{\Delta}~,
\end{equation}
where the arguments of $P$ denote integral GNO flux (restricted to a Weyl chamber of the weight lattice of the GNO dual gauge group as described in \cite{Cremonesi:2013lqa}), $\zeta_{A,B,C}$ are fugacities for the $U(1)^3$ topological symmetry, $\Delta$ is a monopole scaling dimension for operators charged under the GNO flux, and
\begin{eqnarray}\label{PHLii}
P(a_{A,1}=a_{A,1},a_{B,1}=a_{B,2},a_{C,1}=a_{C,2})&=&{1\over(1-t^2)^3}~, \cr P(a_{A,1}>a_{A,1},a_{B,1}=a_{B,2},a_{C,1}=a_{C,2})&=&P(a_{A,1}=a_{A,1},a_{B,1}>a_{B,2},a_{C,1}=a_{C,2})=\cr P(a_{A,1}=a_{A,1},a_{B,1}=a_{B,2},a_{C,1}>a_{C,2})&=&{1\over(1-t)(1-t^2)^2}~,\cr P(a_{A,1}>a_{A,1},a_{B,1}>a_{B,2},a_{C,1}=a_{C,2})&=&P(a_{A,1}>a_{A,1},a_{B,1}=a_{B,2},a_{C,1}>a_{C,2})=\cr P(a_{A,1}=a_{A,1},a_{B,1}>a_{B,2},a_{C,1}>a_{C,2})&=&{1\over(1-t)^2(1-t^2)}~,\cr P(a_{A,1}>a_{A,1},a_{B,1}>a_{B,2},a_{C,1}>a_{C,2})&=&{1\over(1-t)^3}~.
\end{eqnarray}
The monopole scaling dimension in \eqref{3DT332} is given by \cite{Buican:2014hfa}
\begin{eqnarray}\label{monodim}
\Delta&=&{1\over2}\Big(|a_{A,1}|+|a_{A,2}|\Big)+{1\over2}\Big(|a_{A,1}-a_{B,1}|+|a_{A,2}-a_{B,1}|+|a_{A,1}-a_{B,2}|+|a_{A,2}-a_{B,2}|\cr&+&|a_{A,1}-a_{C,1}|+|a_{A,2}-a_{C,1}|+|a_{A,1}-a_{C,2}|+|a_{A,2}-a_{C,2}|+|a_{B,1}-a_{C,1}|\cr&+&|a_{B,2}-a_{C,1}|+|a_{B,1}-a_{C,2}|+|a_{B,2}-a_{C,2}|\Big)-\Big(|a_{A,1}-a_{A,2}|+|a_{B,1}-a_{B,2}|\cr&+&|a_{C,1}-a_{C,2}|\Big)~.
\end{eqnarray}
After identifying fugacities according to
\begin{equation}\label{fmap2}
\zeta_A=wz_1^{-2}z_2^{-1}~, \ \ \ \zeta_B=z_1z_2^2~, \ \ \ \zeta_C=z_1z_2^{-1}~,
\end{equation}
we can then expand the HL index in $t$ to find
\begin{eqnarray}\label{T332HLpert}
\CI_{HL}^{\CT_{3,{3\over2}}}(t)&=&1+\chi_{1}t^{1\over2}+(2\chi_{2}+\chi_{1,1})t+(\chi_{1}+2\chi_{3}+2\chi_{1}\chi_{1,1})t^{3\over2}+(2+\chi_{2}+3\chi_{4}+\cr&+&2\chi_{1,1}+3\chi_{2}\chi_{1,1}+\chi_{2,2})t^2+(3\chi_{5}+\chi_{3}(2+4\chi_{1,1})+\chi_{1}(2+\chi_{1,1}+\cr&+&\chi_{3,0}+\chi_{0,3}+2\chi_{2,2}))t^{5\over2}+\CO(t^3) 
\end{eqnarray}
We immediately see a free hypermultiplet at $\CO(t^{1\over2})$ as expected from our discussion in the main text. Stripping off this free hypermultiplet, we get the putative HL index of the $\CT_X$ theory
\begin{eqnarray}\label{TXHLpert2}
\CI_{HL}^{\CT_X}(t,w,z_1,z_2)&=&1+(\chi_{2}+\chi_{1,1})t+\chi_{1}\chi_{1,1}t^{3\over2}+(1+\chi_{4}+\chi_{1,1}+\chi_{2}\chi_{1,1}+\chi_{2,2})t^2+\cr&+&(\chi_{1} \chi_{1,1} + \chi_{3}\chi_{1,1}+\chi_{1}\chi_{3,0} + 
\chi_{1} \chi_{0,3} + \chi_{1}\chi_{2, 2})t^{5\over2}+\CO(t^3)~,\ \ \ \ \ \ 
\end{eqnarray}
described around \eqref{TXHLpert}.

\end{appendices}
\newpage
\bibliography{chetdocbib}

\end{document}